\documentclass[twocolumn]{aastex7}

\usepackage{enumitem}
\usepackage{xspace}
\usepackage{multirow}
\usepackage{caption}
\usepackage{amssymb,amsmath,amsthm}

\usepackage{lipsum}

\shorttitle{Resolving Nearby Supermassive Black Holes with the Black Hole Explorer}
\shortauthors{Akiyama et al.}
\graphicspath{{./}{figures/}}

\def\m87{M87$^*$\xspace}
\def\sgra{Sgr~A$^*$\xspace}
\def\uas{${\mu}$as\xspace}
\def\comrade{\texttt{Comrade.jl}\xspace}
\def\ehtim{\texttt{eht-imaging}\xspace}
\def\ngehtsim{\texttt{ngehtsim}\xspace}

\begin{document}

\title{Resolving Nearby Supermassive Black Holes with the Black Hole Explorer}

\newcommand{\yamaguchi}{Graduate School of Sciences and Technology for Innovation, Yamaguchi University, 1677-1, Yoshida, Yamaguchi City, Yamaguchi 753-8512, Japan}
\newcommand{\mithaystack}{Massachusetts Institute of Technology Haystack Observatory, 99 Millstone Rd, Westford, MA 01880, USA}
\newcommand{\heriotwatt}{Institute of Sensors, Signals and Systems, Heriot-Watt University, Edinburgh EH14 4AS, UK}
\newcommand{\mitphysics}{Department of Physics, Massachusetts Institute of Technology, 77 Massachusetts Ave, Cambridge, MA 02139, USA}
\newcommand{\naojmizusawa}{Mizusawa VLBI Observatory, National Astronomical Observatory of Japan, 2-12, Mizusawa Hoshigaoka-cho, Oshu-shi, Iwate 023-0801, Japan}
\newcommand{\harvardbhi}{Black Hole Initiative at Harvard University, 20 Garden St, Cambridge, MA 02138, USA}
\newcommand{\cfa}{Center for Astrophysics $|$ Harvard \& Smithsonian, 60 Garden Street, Cambridge, MA 02138, USA}
\newcommand{\denkidai}{Division of Science, School of Science and Engineering, Tokyo Denki University, Ishizaka, Hatoyama-machi, Hiki-gun, Saitama 350-0394, Japan}
\newcommand{\musashino}{Department of Mathematical Engineering, Faculty of Engineering, Musashino University, 3-3-3 Ariake Koto-ku Tokyo 135-8181, Japan}
\newcommand{\ibaraki}{College of Science, Ibaraki University, 2-1-1 Bunkyo, Mito, Ibaraki 310-8512, Japan}
\newcommand{\finca}{Finnish Centre for Astronomy with ESO, University of Turku, FI-20014 Turku, Finland}
\newcommand{\aalto}{Aalto University Mets\"{a}hovi Radio Observatory, Mets\"{a}hovintie 114, FI-02540 Kylm\"{a}l\"{a}, Finland}
\newcommand{\ucon}{Astronomy Department, Universidad de Concepci\'on, Casilla 160-C, Concepción, Chile}

\correspondingauthor{Yuto Akiyama}

\author[orcid=0009-0007-5991-9854, gname=Yuto, sname=Akiyama]{Yuto Akiyama}
\affiliation{\yamaguchi}
\email[show]{f001wbw@yamaguchi-u.ac.jp}

\author[orcid=0000-0002-9475-4254, gname=Kazunori, sname=Akiyama]{Kazunori Akiyama}
\affiliation{\heriotwatt}
\affiliation{\mithaystack}
\affiliation{\naojmizusawa}
\affiliation{\cfa}
\email{k.akiyama@hw.ac.uk}

\author[orcid=0000-0002-5278-9221, gname=Dominic, sname=Pesce]{Dominic W. Pesce}
\affiliation{\cfa}
\affiliation{\harvardbhi}
\email{dpesce@cfa.harvard.edu}

\author[orcid=0000-0002-7179-3816, gname=Daniel, sname=Palimbo]{Daniel C. M. Palumbo}
\affiliation{\cfa}
\affiliation{\harvardbhi}
\email{daniel.palumbo@cfa.harvard.edu}

\author[orcid=0000-0001-5287-0452, gname=Angelo, sname=Ricarte]{Angelo Ricarte}
\affiliation{\cfa}
\affiliation{\harvardbhi}
\email{angelo.ricarte@cfa.harvard.edu}

\author[orcid=0000-0003-3826-5648, gname=Paul, sname=Tiede]{Paul Tiede}
\affiliation{\cfa}
\affiliation{\harvardbhi}
\email{paul.tiede@fas.harvard.edu}

\author[orcid=0009-0003-6164-7406, gname=Marvin, sname=Martinez]{Marvin N. Martinez}
\affiliation{\mitphysics}
\affiliation{\mithaystack}
\email{marvinmn@mit.edu}

\author[orcid=0009-0006-1169-6009, gname=Hikaru, sname=Yoshida]{Hikaru Yoshida}
\affiliation{\ibaraki}
\affiliation{\denkidai}
\email{25nm157h@vc.ibaraki.ac.jp}

\author[orcid=0000-0002-3351-760X, gname=Avery, sname=Broderick]{Avery E. Broderick}
\affiliation{Perimeter Institute for Theoretical Physics, 31 Caroline Street North, Waterloo, ON, N2L 2Y5, Canada}
\affiliation{Waterloo Centre for Astrophysics, University of Waterloo, Waterloo, ON N2L 3G1, Canada}
\affiliation{Department of Physics and Astronomy, University of Waterloo, 200 University Avenue West, Waterloo, ON, N2L 3G1, Canada}
\email{abroderick@perimeterinstitute.ca}

\author[orcid=0000-0002-9221-2910, gname=Aya, sname=Higuchi]{Aya E. Higuchi}
\affiliation{\musashino}
\affiliation{\denkidai}
\email{a-higuchi@musashino-u.ac.jp}

\author[orcid=0000-0002-5297-921X, gname=Sara, sname=Issaoun]{Sara Issaoun}
\affiliation{\cfa}
\affiliation{\harvardbhi}
\email{sara.issaoun@cfa.harvard.edu}

\author[orcid=0000-0001-6920-662X, gname=Neil, sname=Nagar]{Neil M Nagar}
\affiliation{\ucon}
\email{nagar@astro-udec.cl}

\author[orcid=0000-0002-8169-3579, gname=Kotaro, sname=Niinuma]{Kotaro Niinuma}
\affiliation{\yamaguchi}
\email{niinuma@yamaguchi-u.ac.jp}

\author[orcid=0000-0002-9248-086X, gname=Venkatessh, sname=Ramakrishnan]{Venkatessh Ramakrishnan}
\affiliation{\finca}
\affiliation{\aalto}
\email{venkatessh.ramakrishnan@aalto.fi}

\author[orcid=0009-0007-5412-1894, gname=Xinyue, sname=Zhang]{Xinyue Alice Zhang}
\affiliation{\cfa}
\affiliation{\harvardbhi}
\email{xyazhang@stanford.edu}

\received{June 30, 2026}
\revised{August 3, 2026}
\accepted{August 9, 2026}
\submitjournal{Publications of the Astronomical Society of the Pacific}

\begin{abstract}
Recent Event Horizon Telescope results have demonstrated unique and transformative science in gravitational physics and black hole astrophysics enabled by event-horizon-scale imaging of supermassive black holes (SMBHs).
Nevertheless, the angular resolution of current ground-based very long baseline interferometry (VLBI) arrays limits such studies to only two sources, precluding systematic investigations of horizon-scale emission across a nearby SMBH population.
The proposed Black Hole Explorer (BHEX), a millimeter/submillimeter space VLBI mission, would overcome this limitation by delivering substantially higher angular resolution.
Here, we present a series of simulated observations to assess a population of nearby horizon-scale targets accessible with BHEX.
Based on a recently developed SMBH number density model, we find that BHEX could infer black hole masses for $\sim$70--90 sources from size measurements, constrain magnetic field structures through linear polarization imaging for $\sim$20--30 sources, and resolve black hole shadows for $\sim$20--25 sources.
Targeted observations of $\sim$50 nearby SMBHs are expected to yield measurements for $\sim$30 source sizes and $\sim$10 shadows and linear-polarization patterns.
These projections are supported by detailed imaging simulations of general relativistic magnetohydrodynamic (GRMHD) models for eleven nearby SMBHs.
Together, our results highlight BHEX as a powerful facility for revealing the demographics of SMBH properties across diverse accretion states, radio loudness, host galaxy environments, and viewing geometries.

\end{abstract}


\section{Introduction}\label{sec:intro}
The Event Horizon Telescope \citep[EHT;][]{EHTC2017M87Paper2} has opened a fundamentally new observational window by resolving and directly imaging the immediate environment of a black hole event horizon.
Using Earth-sized very long baseline interferometry (VLBI) at 1.3\,mm (or 230\,GHz), the EHT has produced the first spatially resolved images of black hole shadows, revealed as dark central depressions surrounded by bright rings of synchrotron emission from hot, orbiting plasma.
To date, such images have been obtained for two supermassive black holes (SMBHs): Messier~87$^{*}$ (\m87) in the nearby radio galaxy M87 \citep{EHTC2017M87Paper1,EHTC2017M87Paper4,EHTC2017M87Paper7,EHTC2017M87Paper9,EHTC2018M87Paper1,EHTC2021M87Paper1}, and Sagittarius~A$^{*}$ (\sgra) at the center of the Milky Way \citep{EHTC2017SgrAPaper1,EHTC2017SgrAPaper3,EHTC2017SgrAPaper7}.

These horizon-scale images enable new and stringent tests of general relativity and alternative theories of gravity \citep{EHTC2017SgrAPaper6}, as well as direct and accurate measurements of black hole masses \citep{EHTC2017M87Paper6,EHTC2018M87Paper2}.
In addition, polarimetric imaging of the emission rings reveals strongly magnetized accretion flows around both SMBHs \citep{EHTC2017M87Paper8,EHTC2017SgrAPaper8,EHTC2021M87Paper1}, supporting theoretical models in which relativistic jets are powered by efficient extraction of rotational energy from spinning black holes \citep{Blandford_2019}.
Together, these observations provide compelling evidence for the existence of black holes at galactic centers and their role as the engines of relativistic jets.

While these results demonstrate the transformative potential of event-horizon-scale observations, they also highlight fundamental limitations imposed by Earth-based VLBI.
The angular resolution of current ground-based arrays is ultimately constrained by the diameter of the Earth, while severe atmospheric absorption and phase instability limit observations at higher frequencies that would otherwise provide finer resolution \citep{Pesce_2024_ngehtsim}.
As a result, Earth-bound VLBI has to date only produced horizon-scale images for the two most favorable targets, \m87 and \sgra, and population-level studies of black hole shadows and accretion flows in nearby SMBHs are fundamentally limited by the achievable angular resolution \citep[e.g.,][]{Pesce_2021, Pesce_2022, Ramakrishnan_2023, Faggert_2026}.

The Black Hole Explorer \citep[BHEX;][]{Johnson_2024_SPIE,Akiyama_2024_SPIE}\footnote{\url{https://www.blackholeexplorer.org/}, accessed on June 30, 2026} is a next-generation space VLBI mission concept being developed for the Astrophysics Small Explorer (SMEX) program of the National Aeronautics and Space Administration (NASA).
By extending VLBI baselines into space, BHEX is expected to achieve substantially higher angular resolution, enabling demographic studies of event-horizon-scale black hole properties and providing critical insights into the physical processes governing black hole growth and feedback \citep{Zhang_2025}.

In this paper, we present a series of simulated observations designed to assess a population of nearby horizon-scale targets accessible with BHEX.
We describe the mission specifications and the assumed ground array in \autoref{sec:background}.
In \autoref{sec:Populations_of_SMBHs}, we evaluate the population of nearby SMBHs observable with BHEX using SMBH number density models and known source catalogs.
Detailed imaging simulations based on general relativistic magnetohydrodynamic (GRMHD) models are presented in \autoref{sec:Detailed_Imaging_Simulation}.
We summarize our findings and discuss their implications in \autoref{sec:conclusion}.

\section{Background}\label{sec:background}
\subsection{BHEX Mission Specifications}\label{subsec:bhex}
The BHEX mission is designed to launch a submillimeter/millimeter space telescope to conduct space VLBI observations with ground stations \citep{Johnson_2024_SPIE}.
Here, we outline the technical specifications assumed in this work, largely based on the publicly available instrument design specifications \citep[see][for an overview]{Marrone_2024_SPIE, Peretz_2024_SPIE}.

The BHEX satellite is equipped with a simultaneous dual-band, dual-polarization receiving system made of two cryogenic receivers \citep{Tong_2024_SPIE} leveraging a 4.5-K class mechanical cryocooling system \citep{Rana_2024_SPIE}: a single-side-band (SSB) 100\,GHz-band receiver operating at 80--106\,GHz (henceforth, low band) and a double-side-band (DSB) 300\,GHz-band receiver at 240--320\,GHz (henceforth, high band).
Each of the two BHEX receivers will receive radio signals with a total bandwidth of 8\,GHz per polarization, collected with a 3.4\,m diameter antenna\footnote{Based on the BHEX Fact Sheet available on the BHEX Website as of June 30, 2026.} \citep{Sridharan_2024_SPIE}. The signals will be digitally sampled at 1-bit depth with an aggregate data rate of 64\,Gbps \citep{Srinivasan_2024_SPIE}. The digitally sampled data will be transferred to the optical ground stations (OGSs) with the novel broadband laser communication technology at a downlink data rate of up to $\sim 100$\,Gbps \citep{Wang_2024_SPIE}.

With the simultaneous dual-band receiving capability, the mission leverages the technique of multi-frequency phase referencing through frequency phase transfer \citep[FPT; e.g.,][]{Asaki_1996, Asaki_1998, Dodson_2009, Rioja_2011}, successfully demonstrated with the Korean VLBI Network at frequencies between 22, 43, 80, and 128\,GHz \citep[e.g.,][]{Rioja_2015}, and recently between 80 and 230\,GHz with an intercontinental array of three stations participating in EHT \citep{Issaoun_Pesce_2025, Zhao_2025}.
FPT significantly enhances the fringe detection sensitivity of baselines involving ground stations at higher frequencies, which are more severely limited by turbulent atmospheric phase fluctuations.
FPT enables this by extending the signal coherence time with phase solutions derived at lower frequencies, typically with much better sensitivities due to lower receiver temperatures, lower atmospheric opacity, and higher telescope aperture efficiencies.

For the BHEX mission, the high-band observing frequency is chosen to be a factor of three higher than the low-band frequency \citep{Marrone_2024_SPIE, Tong_2024_SPIE} to avoid potential issues of phase ambiguities in FPT caused by a non-integer frequency ratio \citep[e.g.,][]{Dodson_2009, Rioja_2023}.
In this work, we primarily consider observations at the following two frequency settings: 240\,GHz observations using 80--88\,GHz at the low band and 240--248/248--256\,GHz at the high band; and 320\,GHz observations using 98--106\,GHz at the low band and 296--304/312--320\,GHz at the high band.

The BHEX satellite is planned to have a fully-circular, highly-inclined orbit with the orbital height of $\sim 20,000$\,km or equivalently the orbital period of $\sim 12$\,hours, optimized for the observations of \m87 and \sgra \citep[e.g.,][]{Johnson_2024_SPIE, Marrone_2024_SPIE, Issaoun_2024_SPIE}. The orbit provides the maximum baseline length of $\sim 34,000$\,km to the ground stations, achieving $\sim 3$ times longer maximum baseline and consequently $\sim 3$ times finer angular resolution than those of current ground-based millimeter/submillimeter VLBI arrays.
In this work, we adopt a fully-circular polar orbit with the orbital period of 12 hours, the right ascension of the ascending node of $-112.3^\circ$, and the inclination angle of $90^\circ$, which are all broadly consistent with the publicly available specifications.

The BHEX data received at the OGSs will be correlated with data recorded at partner ground observatories.
Each ground observatory receives signals with a total bandwidth of 8\,GHz per polarization in an SSB at the low band and/or 16\,GHz per polarization in two sidebands at the high band, which together effectively cover the observing frequency bands of the BHEX satellite.
Data will be recorded with 2-bit sampling at aggregate data rates of 64\,Gbps and 128\,Gbps at the low and high bands, respectively \citep{Marrone_2024_SPIE}.

\subsection{Ground Observatories}\label{subsec:pgos}

\begin{figure*}
    \centering
    \includegraphics[width=1\textwidth]{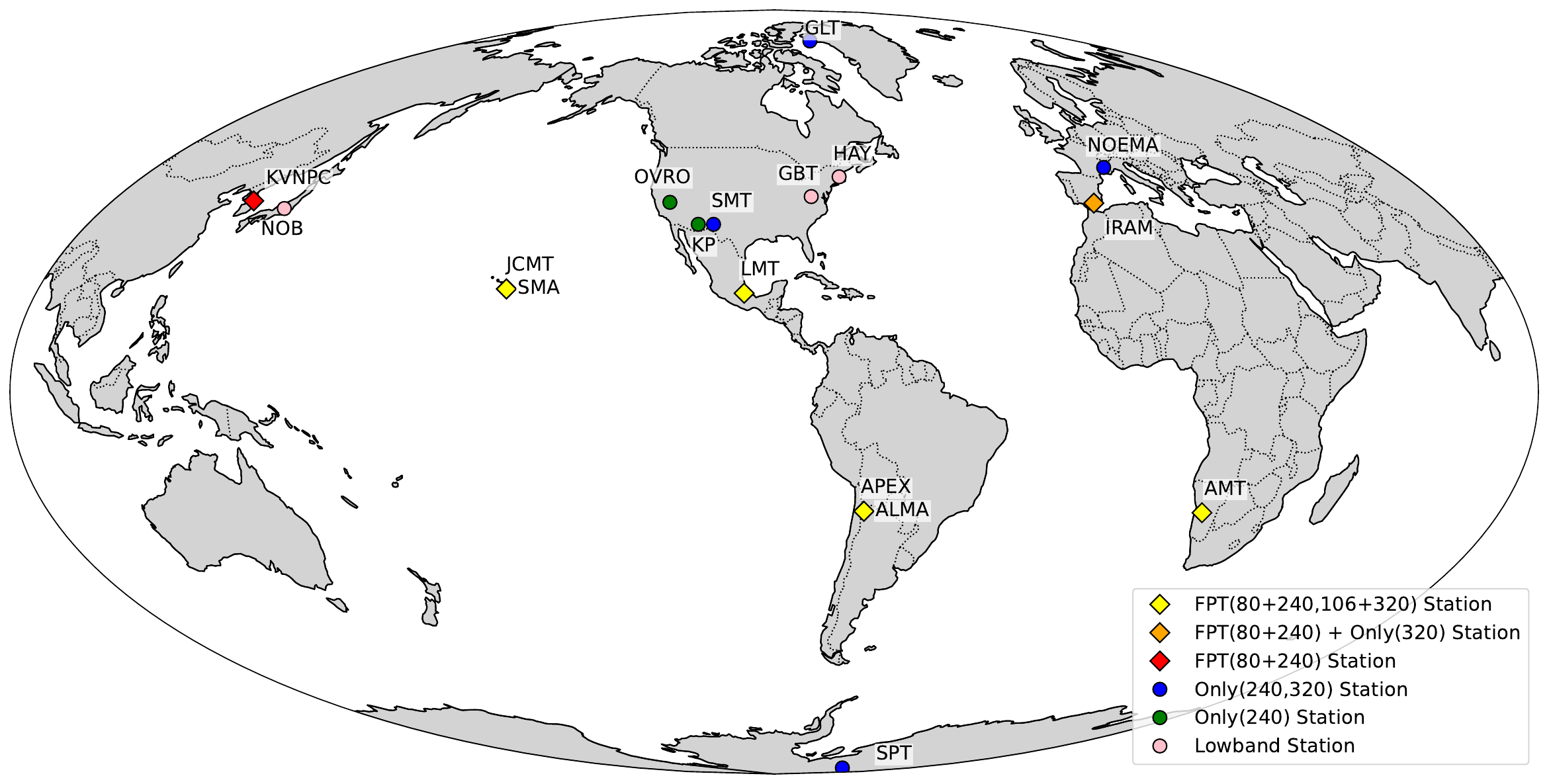}
    \caption{Map of the ground-based stations used in this study and their observing frequencies. Diamond symbols indicate stations capable of FPT, while circular symbols denote stations without FPT capability. The frequencies listed in parentheses indicate the available observing bands at each station. For example, IRAM supports FPT between 80 and 240\,GHz, but FPT is not available at 320\,GHz.
    \label{fig:Array_map}
    } 
\end{figure*}

As of May 2025, the EHT array, which conducts VLBI observations at frequencies covering the BHEX observing bands of 240\,GHz and 320\,GHz, consists of 12 observatories distributed across 10 geographic sites \citep[e.g.,][]{EHTC2017M87Paper2, EHT_MSO_2024}.  
By the early 2030s, when the BHEX mission is planned to be operational, further expansion of the ground network is anticipated.  
According to the EHT Mid-Scale Plan \citep{EHT_MSO_2024}, three new observatories are expected to join the network within the next few years, and the more ambitious ngEHT plan envisions the addition of five more observatories \citep{Doeleman_2023}.

In this work, we assume as ground stations the EHT observatories and their receiver systems that already exist or have secured funding for upgrades \citep{EHTC2017M87Paper2, EHT_MSO_2024, Issaoun_Pesce_2025}, given that the BHEX launch is currently expected around 2032.  
In addition to the EHT observatories, we also consider the Green Bank Telescope operated by the Green Bank Observatory and the Nobeyama 45-m Telescope operated by the National Astronomical Observatory of Japan, both of which have been involved in the BHEX mission development as partners.  
Although these telescopes can currently observe only in the low band, their large apertures and high sensitivities may make them promising anchor stations for fringe detection at the low band, which is crucial for conducting FPT.

\begin{deluxetable*}{llcc}
\digitalasset
\tablewidth{0pt}
\tablecaption{Ground observatories considered in this work \label{tab:sites}}
\tablehead{
\colhead{Name (Shortened Name)} & \colhead{Location} & \colhead{240\,GHz Bands} & \colhead{320\,GHz Bands} 
}
\startdata
Africa Millimeter Telescope (AMT)&Gamsberg, Namibia&80+240&106+320\\
Atacama Large Millimeter/submillimeter Array (ALMA)&Atacama Desert, Chile&240&320\\
Atacama Pathfinder Experiment (APEX)&Atacama Desert, Chile&80+240&106+320\\
Green Bank 100-m Telescope (GBT)&West Virginia, USA&80&106\\
Greenland Telescope (GLT)&Greenland, Denmark&240&320\\
Haystack 37-m Telescope (HAY)&Massachusetts, USA&80&106\\
IRAM 30-m Telescope (IRAM)&Sierra Nevada, Spain&80+240&320\\
Kitt Peak National Observatory (KP)&Arizona, USA&80 (S)/240 (N)&106\\
Korean VLBI Network Pyeongchang (KVNPC)&Pyeongchang, Korea&80+240&---\\
Korean VLBI Network Yonsei (KVNYS)&Seoul, Korea&80+240&---\\
Large Millimeter Telescope (LMT)&Sierra Negra, Mexico&80+240&106+320\\
Nobeyama 45-m Telescope (NOB)&Nagano, Japan&80&106\\
Northern Extended Millimeter Array (NOEMA)&Plateau de Bure, France&240&320\\
Owens Valley Radio Observatory (OVRO)&California, USA&240&---\\
Submillimeter Array (SMA) \& & \multirow{2}{*}{Hawaii, USA}&\multirow{2}{*}{80+240}&\multirow{2}{*}{106+320} \\
James Clerk Maxwell Telescope (JCMT)  & \\
Submillimeter Telescope (SMT)&Arizona, USA&240&320\\
South Pole Telescope (SPT)&South Pole, Antarctica&240&320\\
\enddata
\tablecomments{
80 (S)/240 (N) indicates that the Kitt Peak station is assumed to observe at 80\,GHz in the southern campaign and at 240\,GHz in the northern campaign.
}
\end{deluxetable*}

\autoref{tab:sites} lists the ground stations considered in this work, and their geographic distributions are shown in \autoref{fig:Array_map}.  
At the 240\,GHz band, we use seven stations capable of FPT and six stations without FPT capability; at the 320\,GHz band, five stations capable of FPT and five stations without FPT capability are included.

Ground observatories operating at the BHEX frequency bands of 240--320\,GHz are strongly affected by weather conditions.  
Therefore, observations are typically feasible only during the autumn and winter seasons \citep{Raymond_2021}.  
In addition, constraints such as solar avoidance for the BHEX satellite limit the observability of each target to approximately half the year, when its angular separation from the Sun exceeds about $90^\circ$.  
For instance, \cite{Issaoun_2024_SPIE} discusses an operational plan in which the main photon ring targets for BHEX, \m87 and \sgra, are observed at different times: \m87 during the northern hemisphere winter observing campaign (January--March; hereafter, the northern campaign) and \sgra during the southern hemisphere winter observing campaign (July--September; hereafter, the southern campaign).

\section{Populations of SMBHs in Scope}\label{sec:Populations_of_SMBHs}
The VLBI network including BHEX (\autoref{tab:sites}) is expected to achieve a larger maximum baseline than the EHT, thereby filling the outer regions of the $(u,v)$ coverage. This expanded baseline coverage improves the angular resolution, enabling observations of sources whose apparent sizes were previously too small for black hole shadow detection. Consequently, the number of detectable black hole shadows is expected to increase. In this section, following the methodology of \citet{Pesce_2022}, we define the detection thresholds for the black hole mass, magnetic-field, and shadow for SMBHs observed with the VLBI network considered in this work. We then evaluate how many of the currently observed SMBHs in the Event Horizon and Environs (ETHER) database \citep{Ramakrishnan_2023} exceed these thresholds. Next, we assume that uncertainties are present in SMBH mass estimates and model these uncertainties with a lognormal distribution. Based on this model, we perform Monte Carlo simulations that account for the resulting statistical scatter in the shadow diameter and estimate the probability distributions of the number of detectable sources as a function of the number of observed targets.

\subsection{Measurement Thresholds}\label{subsec:measurement_thresholds}

\begin{figure*}[t]
    \centering
    \begin{minipage}{0.34\textwidth}
        \centering
        \includegraphics[width=\linewidth]{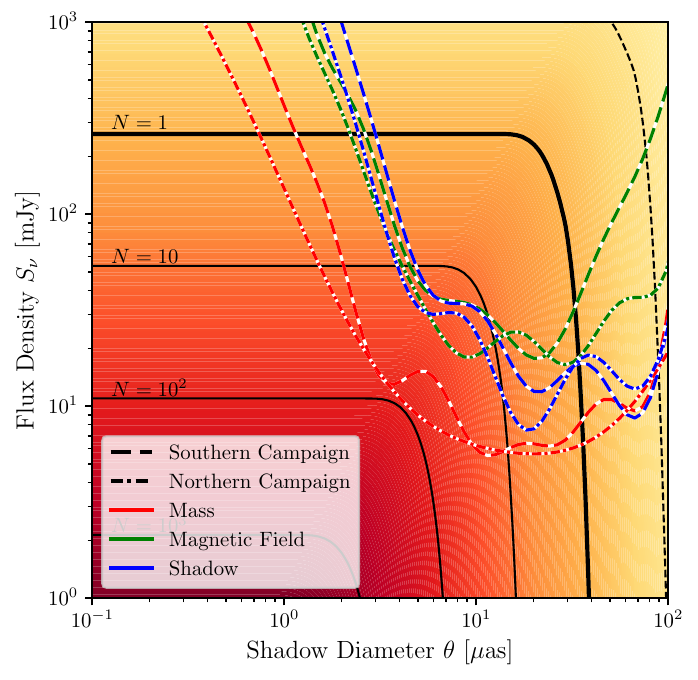}
    \end{minipage}\hfill
    \begin{minipage}{0.33\textwidth}
        \centering
        \includegraphics[width=\linewidth]{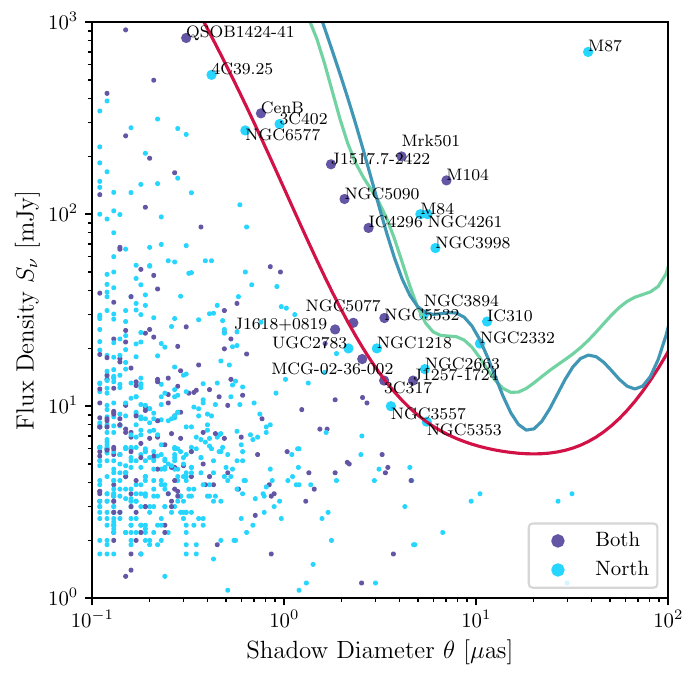}
    \end{minipage}\hfill
    \begin{minipage}{0.33\textwidth}
        \centering
        \includegraphics[width=\linewidth]{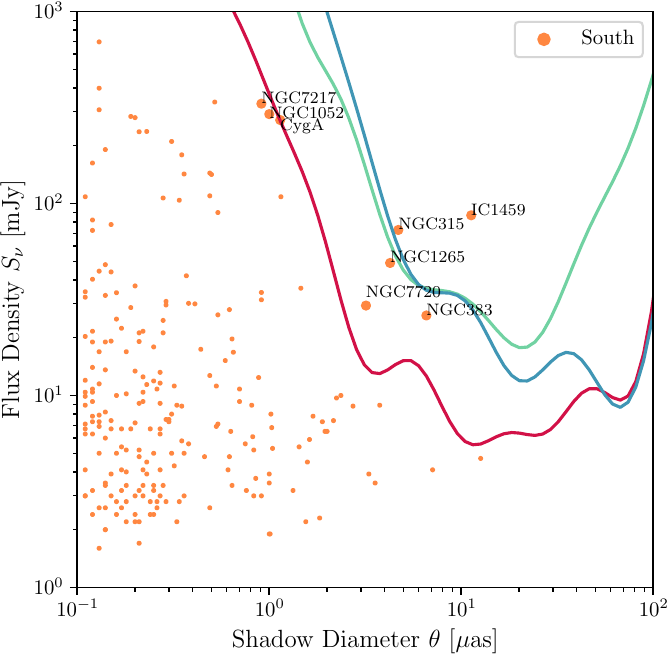}
    \end{minipage}
    \caption{
    Figure showing the detectable threshold lines for each parameter at 240\,GHz, together with the cumulative number density of SMBHs (left) and the ETHER sample in the northern campaign (center) and southern campaign (right).
    In the left panel, the black contour lines indicate the parameter-space regions within which the expected cumulative numbers of SMBHs are 1, 10, and 100.
    The red, green, and blue curves represent the detection threshold lines for the black hole mass, magnetic-field structure, and black hole shadow, respectively.
    In the ETHER sample panels, blue points indicate sources observable during the northern campaign, orange points indicate sources observable during the southern campaign, and purple points denote sources observable during either observing season.
    }
    \label{fig:measurement_thresholds}
\end{figure*}

At the observing frequencies of BHEX, synchrotron emission from nearby SMBHs is expected to be optically thin, allowing the black hole shadow and the surrounding ring-like emission structure to be resolved \citep[e.g.,][]{Zhang_2025}, similar to those observed in EHT observations of \m87\ and \sgra. 
Here, following the methodology of \cite{Pesce_2022}, we define the detection thresholds for the black hole mass, magnetic-field, and shadow for SMBHs observed with BHEX and the ground array (\autoref{tab:sites}). 
We adopt the following observables extracted from the ring-like emission structure, as proxies for these three properties \citep[see,][for detailed definitions and thresholds]{Pesce_2022}.

\paragraph{\noindent Proxy for shadow}
If the emission morphology of a source exhibits a ring-like structure together with a central depression, we consider the black hole shadow to be detected.

\paragraph{\noindent Proxy for mass}
The diameter of the black hole shadow is approximately ten times the gravitational radius ($r_g$) \citep{Bardeen_1973}. Accordingly, the angular diameter of the shadow is given by $\theta \approx 10 r_g / D = 10GM / c^2 D$, where $G$, $c$, $D$, and $M$ denote the gravitational constant, the speed of light, the distance to the source, and the black hole mass, respectively. Since the distance $D$ is assumed to be known, the shadow angular diameter $\theta$ and the black hole mass $M$ can be inferred from the measured source size $d$ \citep{EHTC2017M87Paper6, EHTC2017SgrAPaper6}. We therefore adopt the source size, $d$, as a proxy for the black hole mass $M$.

\paragraph{Proxy for magnetic field}
Because synchrotron radiation is linearly polarized perpendicular to the projected magnetic field on the sky, polarized images allow constraints on the accretion flow's magnetic field state and geometry \citep{EHTC2017M87Paper8,EHTC2017SgrAPaper8}. As a proxy for measuring the magnetic field, we consider measurement of the complex $\beta_2$ coefficient introduced in \citep{Palumbo_2020}, which quantifies the pitch angle and strength of the rotationally symmetric mode of the polarization pattern. The magnetic field geometry is known to correlate with the BH spin parameter, with a precise geometric condition required at the horizon \citep{Blandford&Znajek1977,Chael_2023}. In MAD GRMHD simulations, higher spins result in more toroidal magnetic field structures and therefore more radial polarization patterns \citep{Qiu_2023,Emami_2023,Chael_2023}. Following calibration to spins estimated from photon rings, this signal may one day be utilized in concert with other observables accessible to BHEX \citep[e.g., ring asymmetry][]{Medeiros_2022,Bernshteyn_2026} to infer model-dependent spin constraints on the targets explored in this study.

To assess how precisely the proxies %
can be measured, we performed a model-fitting analysis on synthetic VLBI data generated with \ngehtsim \citep{Pesce_2024_ngehtsim} using BHEX and its ground array (\autoref{tab:sites}). As an emission model for SMBHs, we adopted 
a polarized $m$-ring model \citep{Johnson2020mRing}. 
For each log-uniform grid in the parameter space of the shadow angular diameter and the total horizon-scale flux density, 
$(d, S_0)$,
a series of polarized ring images are generated. To produce a diverse set of polarized ring morphologies, the ring asymmetry and its orientation, as well as the polarization parameters, were randomly sampled. In particular, we considered source sizes in the range $d \in [0.1,100]$\,\uas and flux densities $S_0 \in [10^{-3},1]$\,Jy, fixed the FWHM to $W=d/3$, and set the remaining parameters to match those in \cite{Pesce_2022}.
Two representative source locations are considered: \m87\ as a northern target and IC\,1459 as a southern target.

For the three proxies, we define explicit criteria for considering each quantity to be measurable from the observational data. We estimate the precision of these estimates using a Fisher matrix approach implemented within the \texttt{ngEHTforecast} package (see \autoref{app:fisher_information}; \citealt{Pesce_2022} for details). The source size $d$ and $\beta_2$ are considered measurable if their fractional uncertainties are less than 20\% (corresponding to a statistical significance of at least $5\sigma$). The central depression is considered detectable if the fractional ring width $W/d$ deviates significantly from unity and this deviation can be confirmed with an uncertainty of less than 20\%.

The derived threshold curves are shown in \autoref{fig:measurement_thresholds}. 
In this figure, sources that lie above the threshold curve for each of the three proxies are expected to be detectable for the corresponding observable.
The general trends of the threshold lines are similar to those obtained for a ground-only array \citep[see,][]{Pesce_2022}.
The mass, inferred from the size of the source, produces deeper thresholds in both flux density and shadow diameter, since the size of the source can be measured even when it is only marginally resolved \citep[see early EHT experiments; e.g.,][]{Doeleman_2008, Doeleman_2012, Akiyama_2015, Wielgus_2020}. In contrast, the shadow and magnetic field proxies, which are based on the emission properties of the ring, require brighter and larger sources to capture the ring structures.
The threshold lines show a sharp decline at smaller shadow sizes, linked to the instrumental angular resolution, as well as a modest rise at larger shadow sizes, where the emission begins to be over-resolved by the interferometric array.

\begin{figure}[h]
    \centering
    \includegraphics[width=0.33\textwidth]{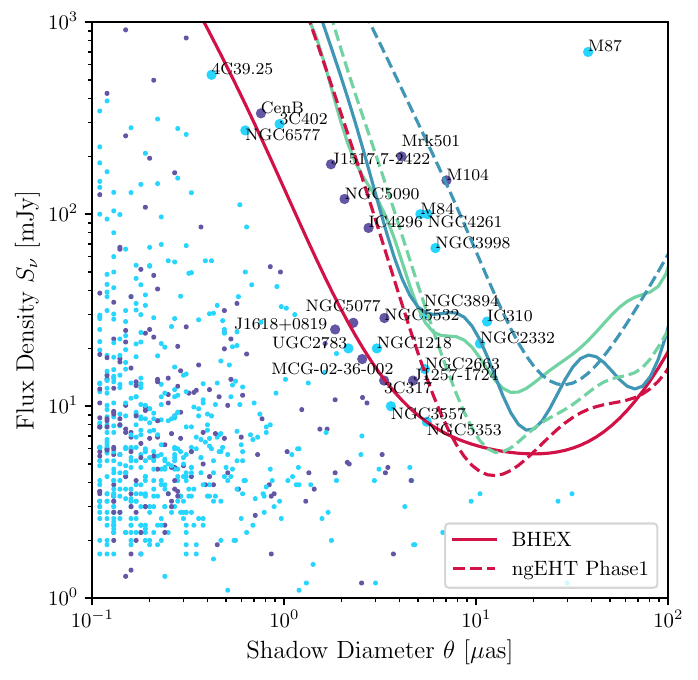}
    \caption{
        Comparison of the detection threshold curves at 240\,GHz for the BHEX array and a ground-only ngEHT Phase 1. Solid lines show the BHEX results obtained in this study, while dashed lines indicate the ground-only case based on the ngEHT Phase 1 presented by \citet{Pesce_2022}.
        }
    \label{fig:BHEX_vs_ngEHT}
\end{figure}

\begin{figure*}[t]
    \centering
    \begin{minipage}{0.33\textwidth}
        \centering
        \includegraphics[width=\linewidth]{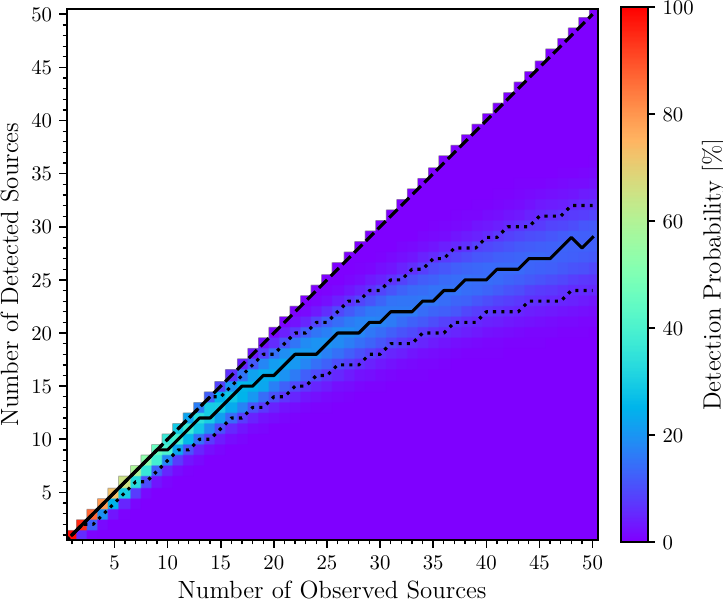}
    \end{minipage}\hfill
    \begin{minipage}{0.33\textwidth}
        \centering
        \includegraphics[width=\linewidth]{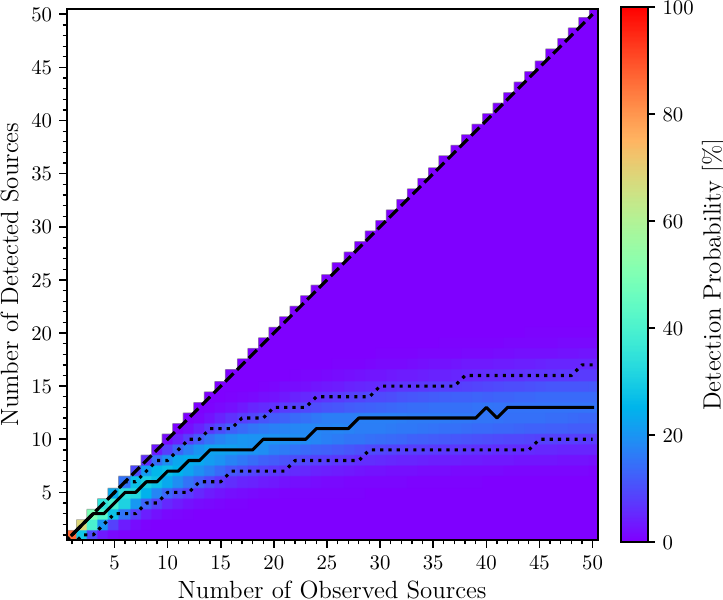}
    \end{minipage}\hfill
    \begin{minipage}{0.33\textwidth}
        \centering
        \includegraphics[width=\linewidth]{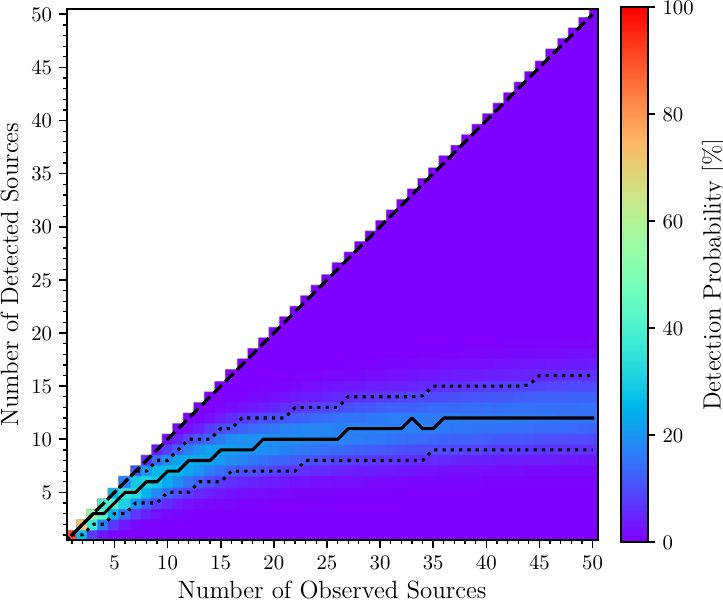}
    \end{minipage}
    \caption{
    Probability distributions of the number of sources for which each observable (left: mass, center: magnetic field, right: shadow) can be detected (vertical axis), estimated for each of the number of the observed sources (horizontal axis). 
    Each of the distributions is estimated for the case to observe sources in descending order of detectability.
    The black solid line indicates the mode of the distribution, while the black dotted lines denote the 10--90\% percentile range. 
    }
    \label{fig:Nobs_vs_Ndet}
\end{figure*}

\autoref{fig:BHEX_vs_ngEHT} compares the detection threshold curves obtained with the BHEX array and a ground-only ngEHT Phase 1 configuration \citep{Pesce_2022}. Compared to a ground-only array, the substantially higher angular resolution provided by BHEX (nominally $\sim 5\,\rm \mu as$) shifts the detection threshold curves toward smaller shadow diameters, thereby increasing the number of detectable sources.

The number of sources exceeding these thresholds can then be estimated based on the cumulative number density model of SMBHs shown in the left panel \citep[see][for details]{Pesce_2021}, yielding 93, 28, and 24 detectable sources for the mass, magnetic field, and shadow proxies, respectively, in the northern campaign, and 71, 19, and 18 in the southern campaign.
The threshold curves for the mass proxy have a characteristic lower limit in compact flux density of $\sim 5$\,mJy, consistent with the anticipated fringe sensitivities for BHEX--ground baselines \citep{Johnson_2024_SPIE}.

\subsection{Measurable ETHER Targets}
Comparison of the detection threshold curves derived in \autoref{subsec:measurement_thresholds} with known samples of SMBHs allows us to infer the list of observable targets, incorporating uncertainties in current measurements.
In the middle and right panels of \autoref{fig:measurement_thresholds}, we show the distribution of sources in the ETHER database along with the threshold lines.
The parameters of ETHER samples are affected by uncertainties in the following quantities: the VLBI-scale compact flux density ($S_\nu$), the black hole mass ($M$), and the distance ($D$), where the latter two provide the estimate of the shadow angular diameter.

The largest uncertainties are anticipated to arise from estimates of the black hole mass, typically inferred from the mass-to-distance ratio, which is proportional to the shadow angular diameter. These estimates are known to suffer from systematic uncertainties of $\sim 0.5$\,dex (about a factor of $\sim 3$), arising from differences among measurement techniques such as stellar dynamics, gas dynamics, and maser observations \citep[e.g.,][]{Pilawa_2025}. This level of uncertainty is consistent with the intrinsic scatter observed in the $M$--$\sigma$ relation \citep[e.g.,][]{Kormendy_and_Ho_2013,Bosch_2016,Saglia_2016}.
This dominates typical uncertainties of $\lesssim 10$\% in distance measurements of the nearby galaxy sample primarily targeted by ETHER \citep[e.g.,][]{Tully_2009,2001Tonry} and $\sim 10$\% uncertainties in compact total flux density estimated from preliminary EHT results for the bright, larger-size end of the ETHER sample.

We estimated the number of ETHER sources for which each of the three proxies would be detectable using Monte Carlo simulations. We first selected 625 ETHER sources with flux densities in the range $1$--$1000$\,mJy and prior estimates of the shadow angular diameter larger than 0.1\,\uas.
Second, ETHER samples are randomly resampled with perturbations in the shadow angular diameter and compared to the threshold lines of the three proxies.
We assume that the shadow angular diameter measurements are subject to a $1\sigma$ systematic uncertainty of 0.5\,dex that follows a lognormal distribution.
By repeating this resampling procedure $10^4$ times, a probability distribution for the number of detectable sources is derived for each proxy.

With a planned mission life of $\sim 2$ years and an aggregate observing time window of $\sim 1$ year with limited duty cycles \citep{Issaoun_2024_SPIE,Johnson_2024_SPIE}, only a fraction of such ETHER samples may be observed by BHEX or any similar ground--space VLBI observatory, given the strong seasonality of ground-based observing.
We therefore derive probabilistic estimates of the anticipated number of detections for each of the three proxies as a function of the number of observed sources, assuming that ETHER targets are observed in order of decreasing detection probability for each proxy.

The results are shown in \autoref{fig:Nobs_vs_Ndet}.
When the top 50 sources from the ETHER database are observed for each proxy, the probability distribution for the number of detections peaks at 29, 13, and 12 for the mass, magnetic field, and shadow proxies, respectively.
The detections of the mass proxy, based on the relatively easy size measurement, increase almost linearly for the first $\sim 20$ observed sources and anticipate a new detection for every $2$--$3$ additional targets up to 50 observed sources.
The other two proxies saturate earlier, as they require larger and brighter targets.

\section{Detailed Imaging Simulations}\label{sec:Detailed_Imaging_Simulation}
\begin{figure*}[t]
    \centering
    
    \begin{minipage}{0.48\linewidth}
        \centering
        \includegraphics[width=\linewidth]{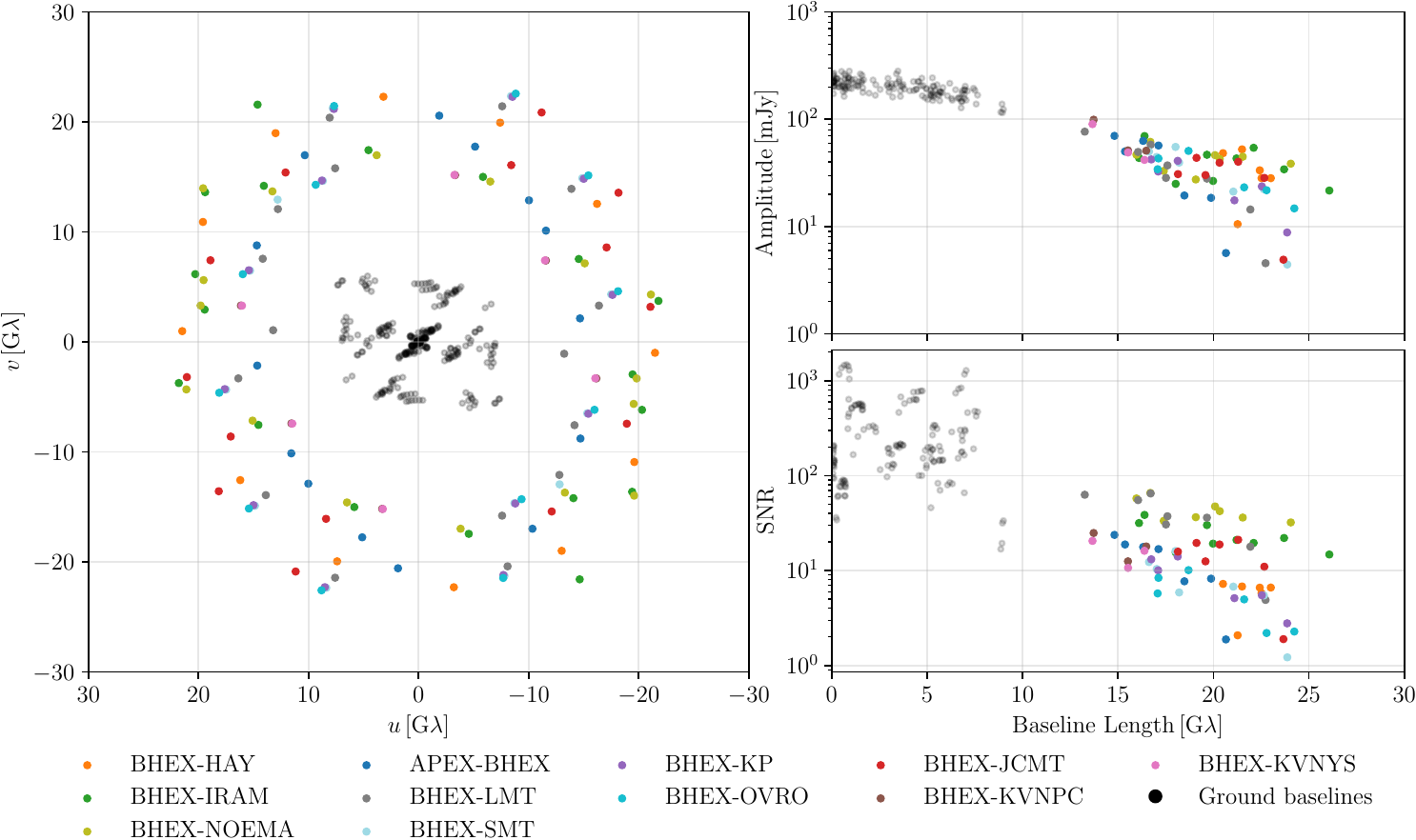}
    \end{minipage}\hfill
    \begin{minipage}{0.48\linewidth}
        \centering
        \includegraphics[width=\linewidth]{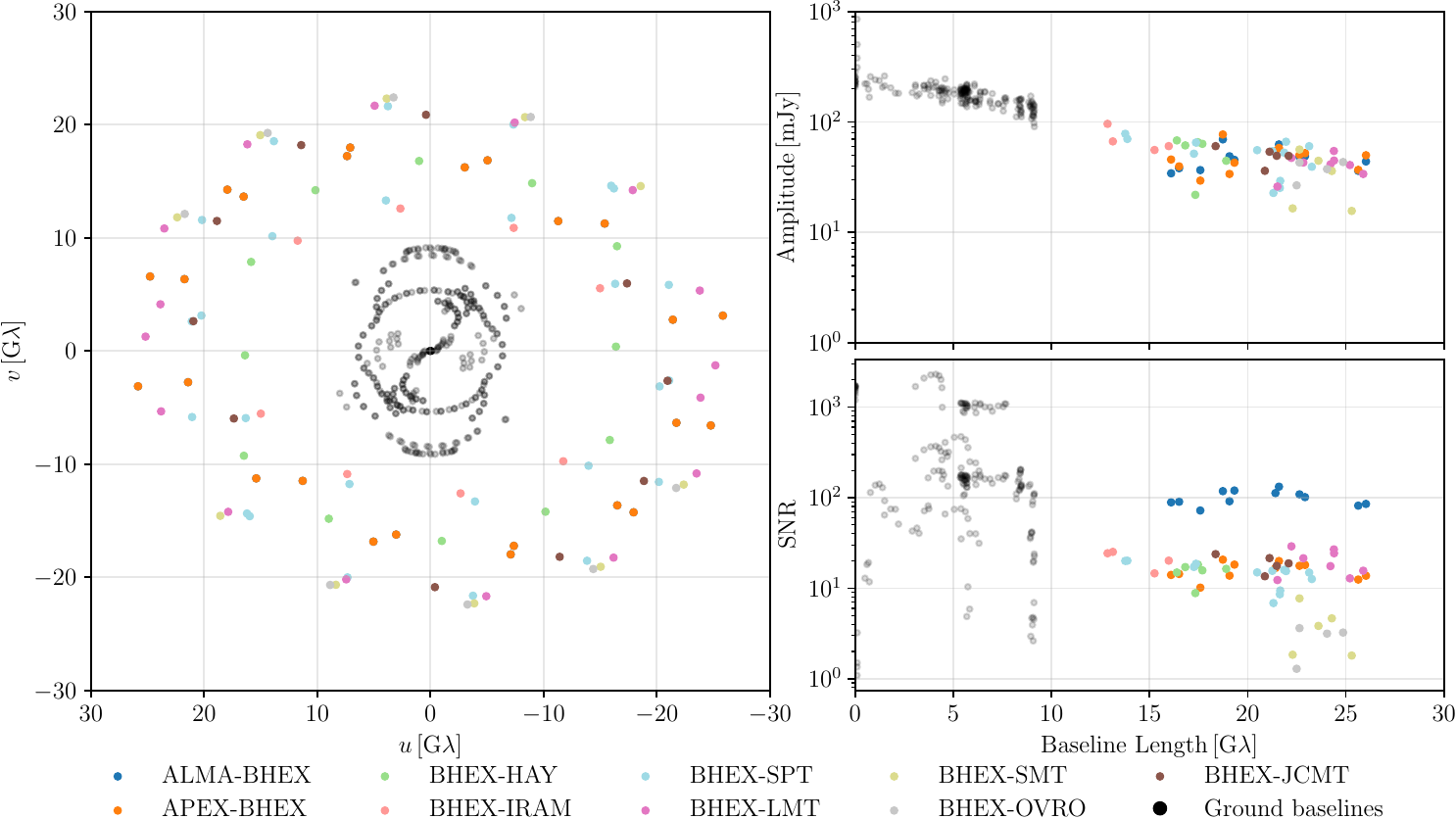}
    \end{minipage}
    \caption{
    Example $(u,v)$ coverage, visibility amplitudes, and signal-to-noise ratios for two representative sources: NGC\,4594 (left) and IC\,1459 (right). Baselines involving BHEX are shown in color, while ground-only baselines are shown in gray.
    }
\label{fig:uv_amp_snr_two_sources}
\end{figure*}

In \autoref{sec:Populations_of_SMBHs}, we estimate the number of sources for which the three proxies can be detected with BHEX using an exhaustive approach with synthetic observations of a geometric  $m$-ring model over a wide range of the source size and brightness. In this section, we examine whether these three proxies remain detectable for more complex GRMHD models. Specifically, we perform synthetic observations of GRMHD models for some ETHER samples and evaluate the performance through the reconstructed images and posterior of emission properties. 

\begin{deluxetable*}{cccccccccc}[t!]
\digitalasset
\tablewidth{0pt}
\tablecaption{The basic properties of the candidate sources\label{tab:shadow_sources}}
\tablehead{
\colhead{Name} & 
\colhead{RA\,[h]} & 
\colhead{Dec\,[$^\circ$]} & 
\colhead{$D\,[\mathrm{Mpc}]$} & 
\colhead{$\log(M_{\bullet}/M_\odot)$} & 
\colhead{Shadow Size\,[${\mu}$as]} & 
\colhead{$S_\nu\,[\mathrm{Jy}]$} & 
\colhead{PA\,[$^\circ$]} & 
\colhead{Observable Period}
}
\startdata
IC1459   & 23.0 & -36.5 & $29 \pm 3.7$   & $9.4^{+0.077}_{-0.035}$ & $8.9$ & $0.22 \pm 0.021$  & $-30$   & May -- Nov \\
NGC4594  & 12.7 & -11.6 & $9.9 \pm 0.82$ & $8.8^{+0.025}_{-0.027}$ & $6.5$ & $0.20 \pm 0.01$   & $-23$   & Jan -- July \\
NGC3998  & 12.0 & 55.5  & $14 \pm 1.3$   & $8.9^{+0.035}_{-0.035}$ & $5.8$ & $0.13 \pm 0.024$  & $-1$    & Nov -- May \\
NGC2663  & 8.8  & -33.8 & $28 \pm 1.4$   & $9.2^{+0.58}_{-0.61}$   & $5.8$ & $0.084 \pm 0.0084$& $-24$   & Nov -- May \\
M84      & 12.4 & 12.9  & $19 \pm 0.60$  & $9.0^{+0.042}_{-0.043}$ & $5.4$ & $0.13 \pm 0.015$  & $1$     & Dec -- June \\
NGC4261  & 12.1 & 5.8   & $31 \pm 1.6$   & $9.2^{+0.11}_{-0.099}$  & $5.2$ & $0.20 \pm 0.05$   & $-91$   & Dec -- June \\
NGC3894  & 11.8 & 59.5  & $50 \pm 2.5$   & $9.4^{+0.48}_{-0.48}$   & $5.2$ & $0.058 \pm 0.015$ & $137.5$ & Nov -- May \\
NGC4552  & 12.6 & 12.6  & $15 \pm 0.99$  & $8.7^{+0.051}_{-0.051}$ & $3.4$ & $0.027 \pm 0.009$ & $49$    & Dec -- June \\
NGC315   & 1.0  & 14.5  & $70 \pm 3.5$   & $9.3^{+0.064}_{-0.033}$ & $2.9$ & $0.18 \pm 0.009$  & $-48$   & July -- Jan \\
NGC1218  & 3.1  & 4.1   & $120 \pm 5.9$  & $9.5^{+0.48}_{-0.48}$   & $2.7$ & $0.11 \pm 0.01$   & $51$    & Aug -- Feb \\
NGC5077  & 13.3 & -12.7 & $39 \pm 8.4$   & $8.9^{+0.18}_{-0.32}$   & $2.0$ & $0.068 \pm 0.007$ & $14$    & Jan -- July \\
\enddata
\tablecomments{
The shadow sizes were computed from the black hole mass $M_{\bullet}$ and distance $D$ following the prescription of \citet{Zhang_2025}. The quantity $S_\nu$ denotes the millimeter-wave flux density. The observable period accounts for the BHEX solar-avoidance constraint, defined as requiring a solar elongation greater than $90^\circ$, thereby preventing direct sunlight from illuminating the spacecraft’s reflective surfaces.
}
\end{deluxetable*}

\subsection{Synthetic Data Generation and Imaging}\label{subsec:Synthetic Data Generation and Imaging}
We simulate synthetic BHEX observations at 240\,GHz and 320\,GHz using \texttt{ngehtsim} \citep{Pesce_2024_ngehtsim}, including thermal noise and complex gain fluctuations in both amplitude and phase.
The synthetic observations reflect realistic weather conditions at each ground observatory site, based on more than a decade of MERRA-2 satellite data, corresponding to the top 15.87\% (i.e., $1\sigma$ equivalent; implemented as the \texttt{good} weather condition).

Source images are taken from GRMHD simulations \citep{Zhang_2025} at an inclination angle of $i = 50^\circ$ for all sources. Each GRMHD image is rotated to roughly match the jet orientation observed at lower frequencies.
The basic properties of the sources and assumed position angles are summarized in \autoref{tab:shadow_sources}.
For IC\,1459, NGC\,315, and NGC\,1218, simulated observations are conducted during the southern campaign window (July--September), while the remaining eight sources are simulated for the northern campaign window (January--March). For each source, nine synthetic observations are performed at intervals of 10 days,  %
yielding a total of 18 synthetic datasets per source across two bands.

To illustrate the observational characteristics of the BHEX array, \autoref{fig:uv_amp_snr_two_sources} presents example $(u,v)$ coverage, visibility amplitudes, and signal-to-noise ratio (SNR) distributions for two representative sources: NGC\,4594 and IC\,1459. The inclusion of BHEX substantially extends the baseline lengths, reaching spatial frequencies of $\sim 25$--$30~\mathrm{G}\lambda$, significantly beyond those achievable with the ground array alone. This extended $(u,v)$ coverage is essential for resolving compact horizon-scale structures.

Although the visibility amplitudes decrease toward longer baselines, as expected for resolved emission, BHEX baselines still retain measurable correlated flux densities at the $\sim 10$--$100~\mathrm{mJy}$ level for both sources. Furthermore, the corresponding SNR distributions indicate that many of these long-baseline detections remain statistically significant, with typical values of $\mathrm{SNR} \sim 5$--$30$. These results demonstrate that BHEX provides both the angular resolution and sensitivity required to probe fine-scale structures associated with the shadow and polarization proxies discussed in this work.

Full polarization images are reconstructed from synthetic BHEX data using \texttt{BHEXScripts.jl}\footnote{\url{https://github.com/ptiede/BHEXScripts.jl}}, an imaging pipeline based on the advanced Bayesian imaging software \texttt{Comrade.jl} \citep{Tiede_2022}.
To evaluate the extraction of the ring-emission properties, we adopt a \texttt{ring} sky model comprising a geometric polarized ring component and a Gaussian Markov random field model for the rest of the emission \citep{Tiede_2025}.
Full posteriors of the sky model and complex gains are sampled using an advanced Hamiltonian Monte Carlo method to evaluate the uncertainties in the ring emission properties.
In this work, each frequency band is reconstructed independently. Incorporating multi-frequency synthesis could further improve the effective $(u,v)$ coverage and imaging fidelity, while also enabling reconstruction of spectral-index and rotation-measure maps; we leave such analyses for future work.

\subsection{Results and Discussions}
\begin{figure*}[t]

\begin{tabular}{cc}
\centering
\begin{minipage}{0.4\textwidth}
    \centering
    \includegraphics[width=\linewidth]{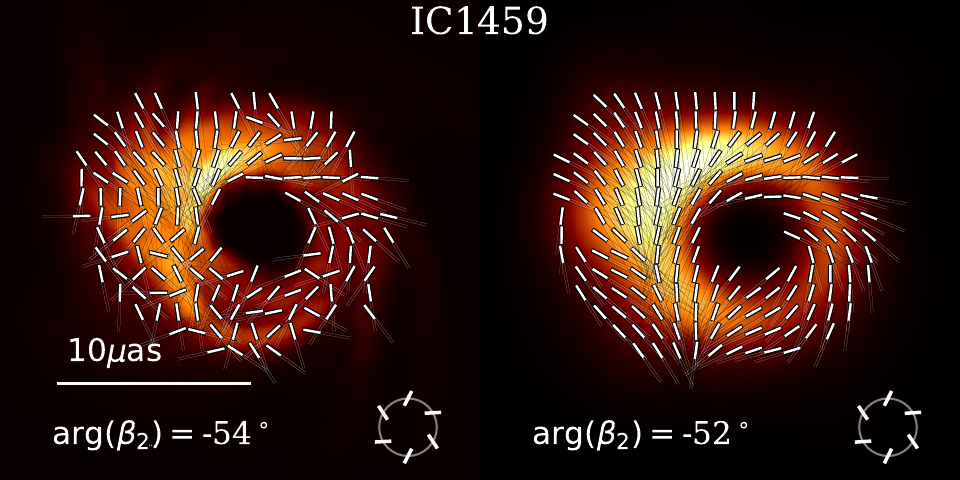}
\end{minipage}
\begin{minipage}{0.4\textwidth}
    \centering
    \includegraphics[width=\linewidth]{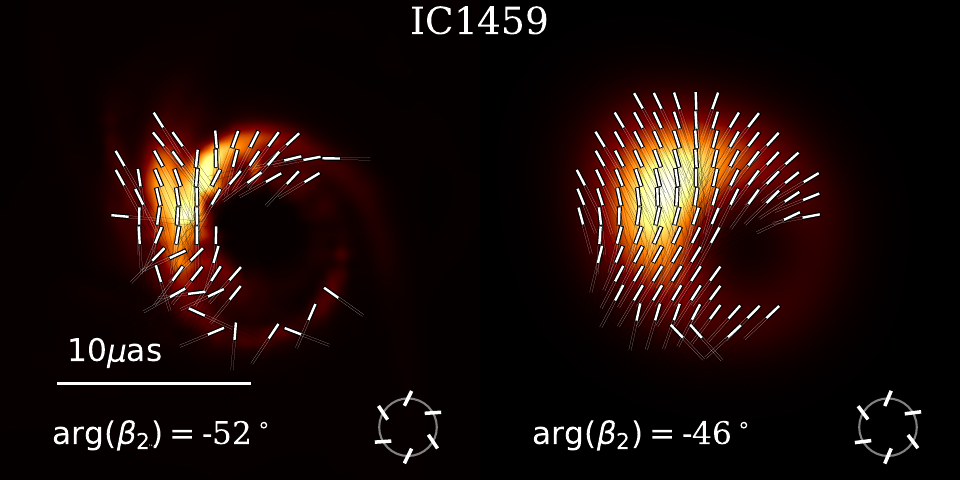}
\end{minipage}
\end{tabular}

\vspace{-0.08em}

\begin{tabular}{cc}
\centering
\begin{minipage}{0.4\textwidth}
    \centering
    \includegraphics[width=\linewidth]{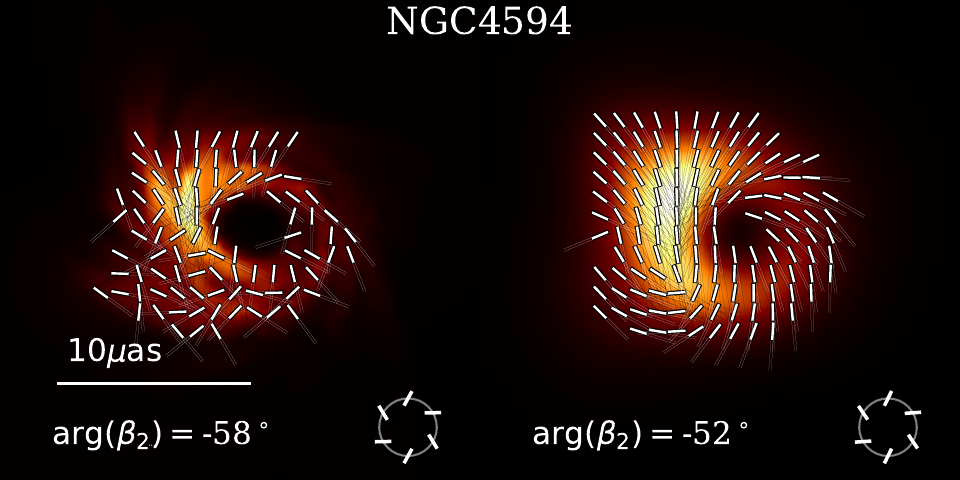}
\end{minipage}
\begin{minipage}{0.4\textwidth}
    \centering
    \includegraphics[width=\linewidth]{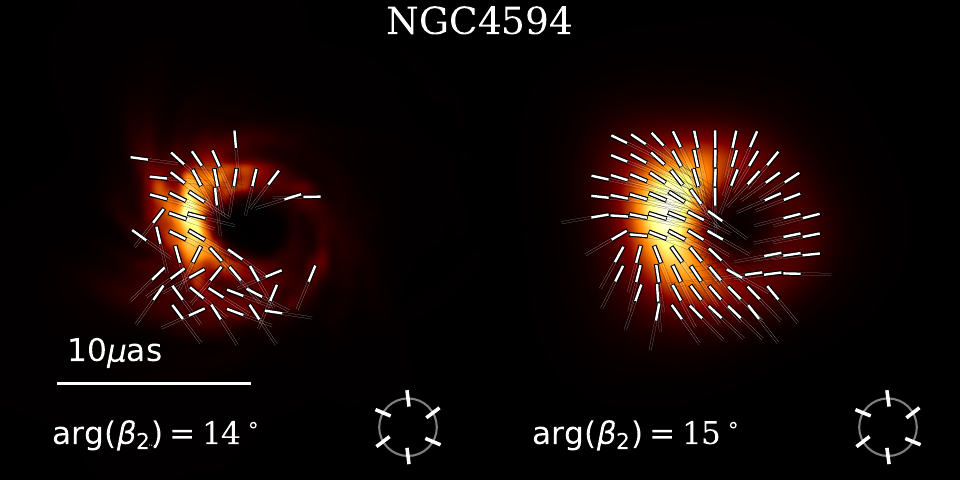}
\end{minipage}
\end{tabular}

\vspace{-0.08em}

\begin{tabular}{cc}
\centering
\begin{minipage}{0.4\textwidth}
    \centering
    \includegraphics[width=\linewidth]{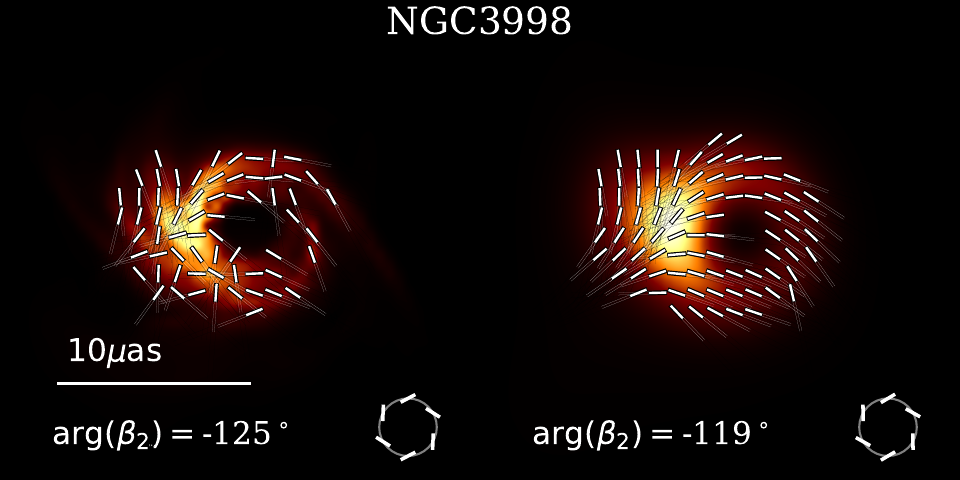}
\end{minipage}
\begin{minipage}{0.4\textwidth}
    \centering
    \includegraphics[width=\linewidth]{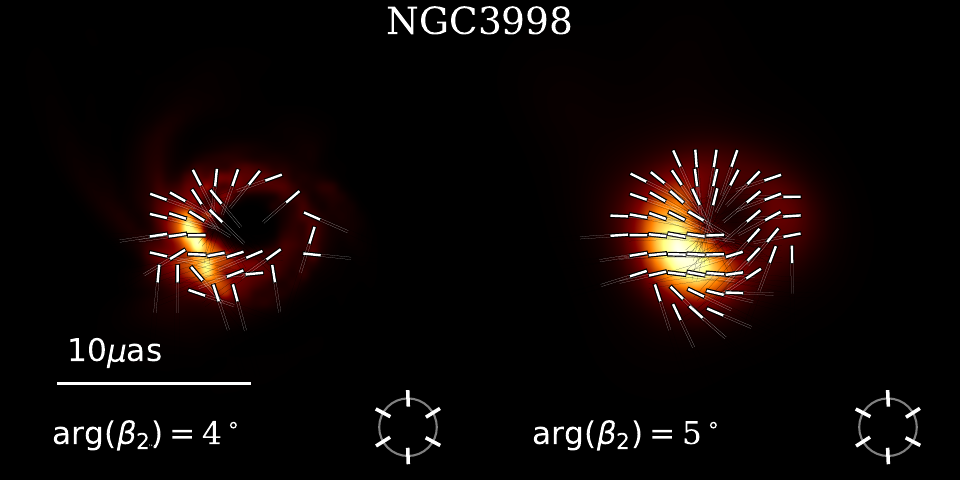}
\end{minipage}
\end{tabular}

\vspace{-0.08em}

\begin{tabular}{cc}
\centering
\begin{minipage}{0.4\textwidth}
    \centering
    \includegraphics[width=\linewidth]{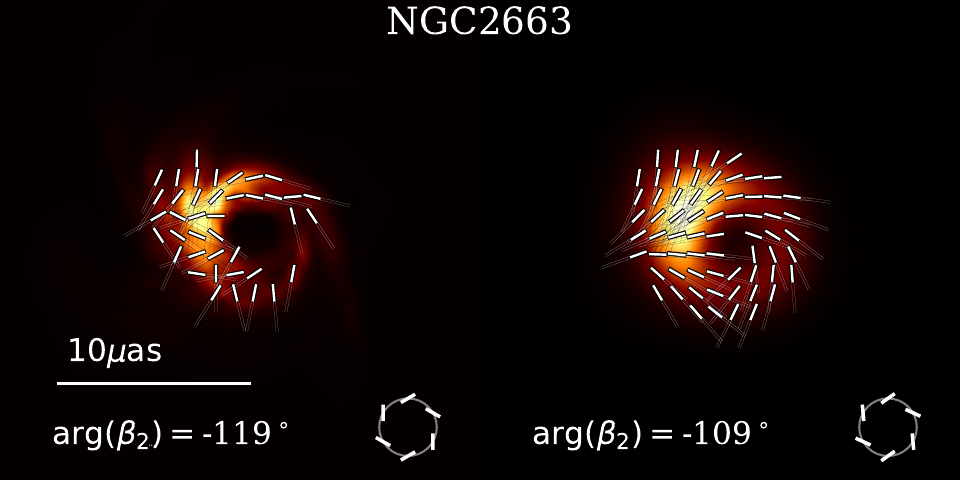}
\end{minipage}
\begin{minipage}{0.4\textwidth}
    \centering
    \includegraphics[width=\linewidth]{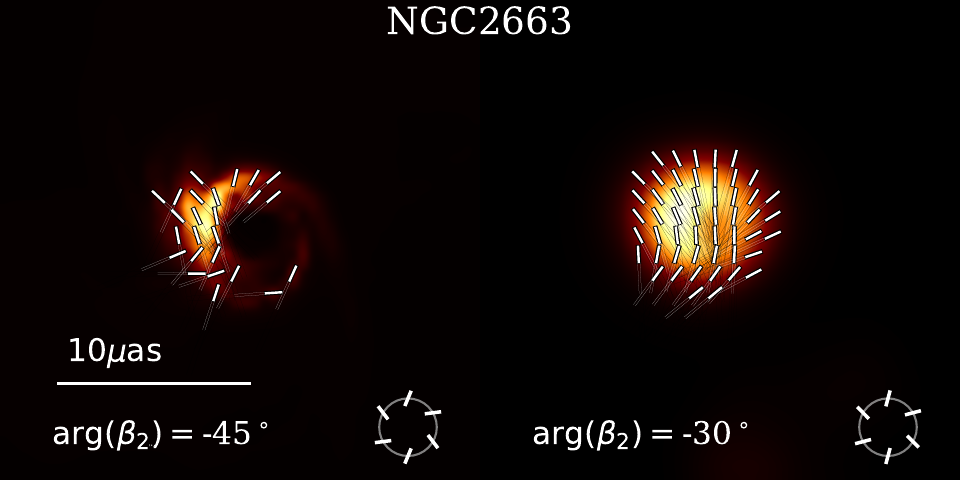}
\end{minipage}
\end{tabular}

\vspace{-0.08em}

\begin{tabular}{cc}
\centering
\begin{minipage}{0.4\textwidth}
    \centering
    \includegraphics[width=\linewidth]{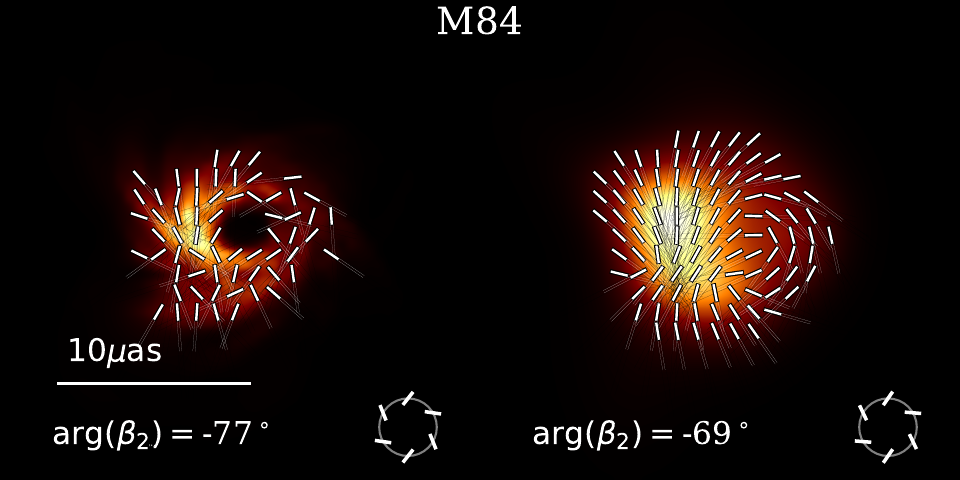}
\end{minipage}
\begin{minipage}{0.4\textwidth}
    \centering
    \includegraphics[width=\linewidth]{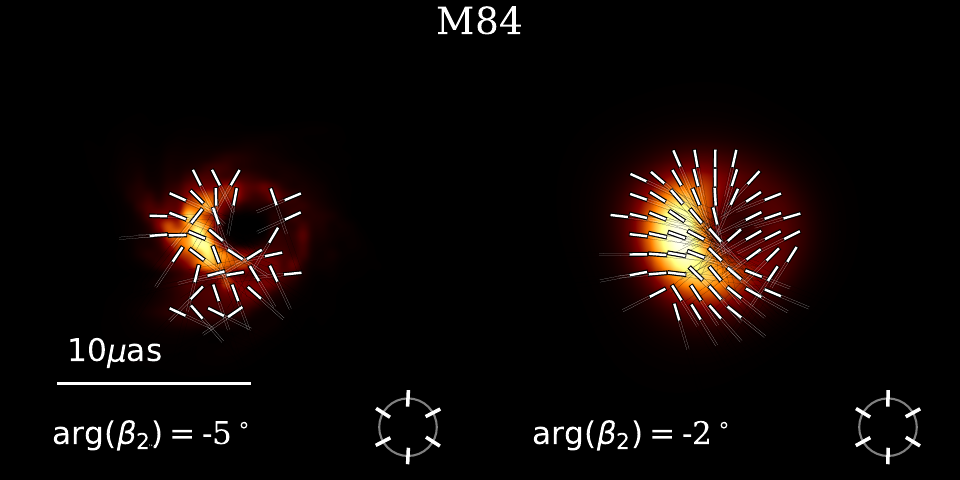}
\end{minipage}
\end{tabular}

\vspace{-0.08em}

\begin{tabular}{cc}
\centering
\begin{minipage}{0.4\textwidth}
    \centering
    \includegraphics[width=\linewidth]{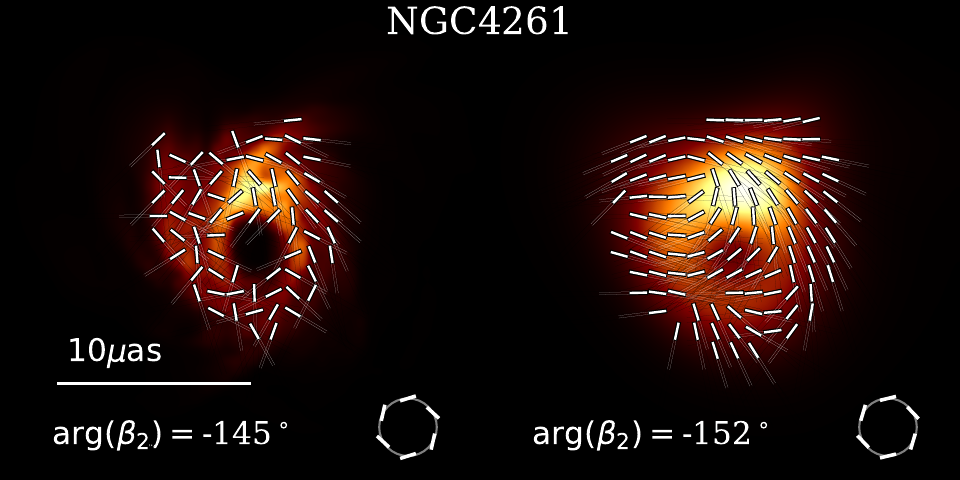}
\end{minipage}
\begin{minipage}{0.4\textwidth}
    \centering
    \includegraphics[width=\linewidth]{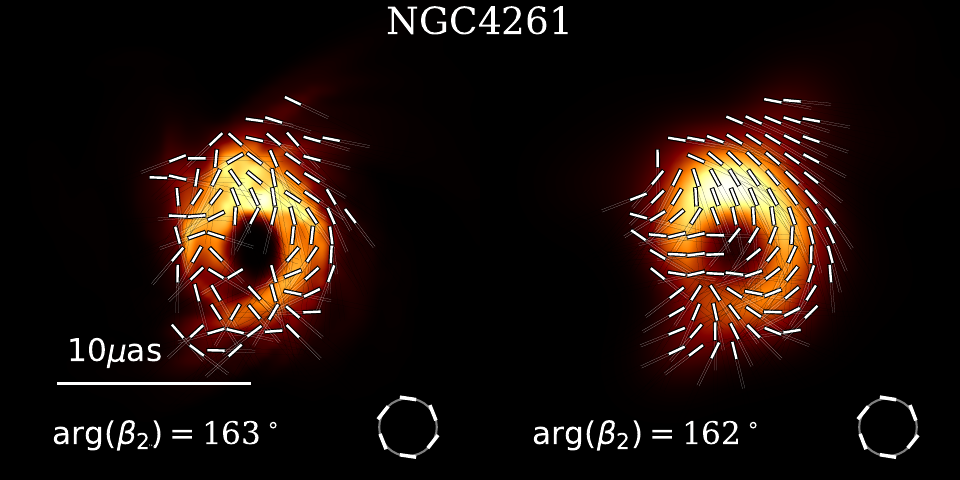}
\end{minipage}

\end{tabular}
\caption{Comparison between the GRMHD model and the reconstructed images. The left panel shows the 240\,GHz observations, and the right panel shows the 320\,GHz observations. Within each panel, the left subpanel displays the GRMHD model and the right subpanel shows the reconstructed image. The $\beta_2$ angle is also indicated for each case.}
\label{fig:images}
\end{figure*}

\setcounter{figure}{5} 

\begin{figure*}
\begin{tabular}{cc}
\centering
\begin{minipage}{0.4\textwidth}
    \centering
    \includegraphics[width=\linewidth]{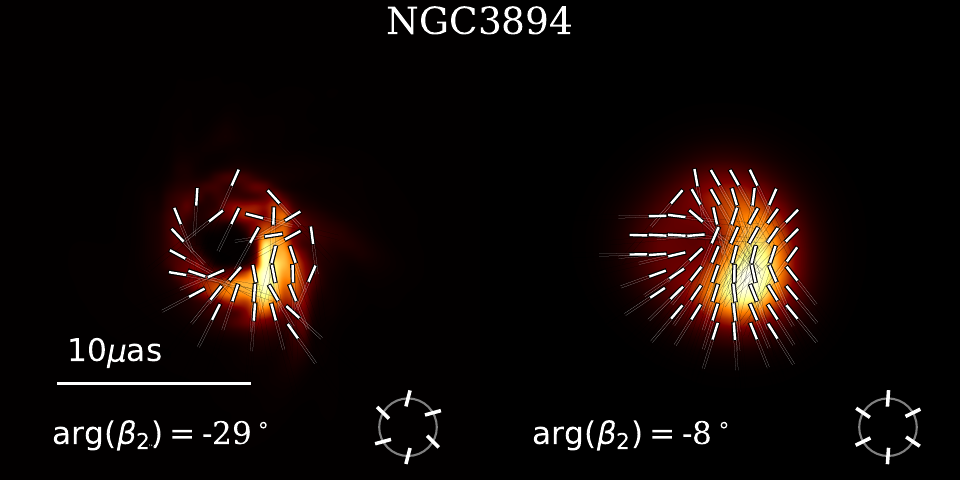}
\end{minipage}
\begin{minipage}{0.4\textwidth}
    \centering
    \includegraphics[width=\linewidth]{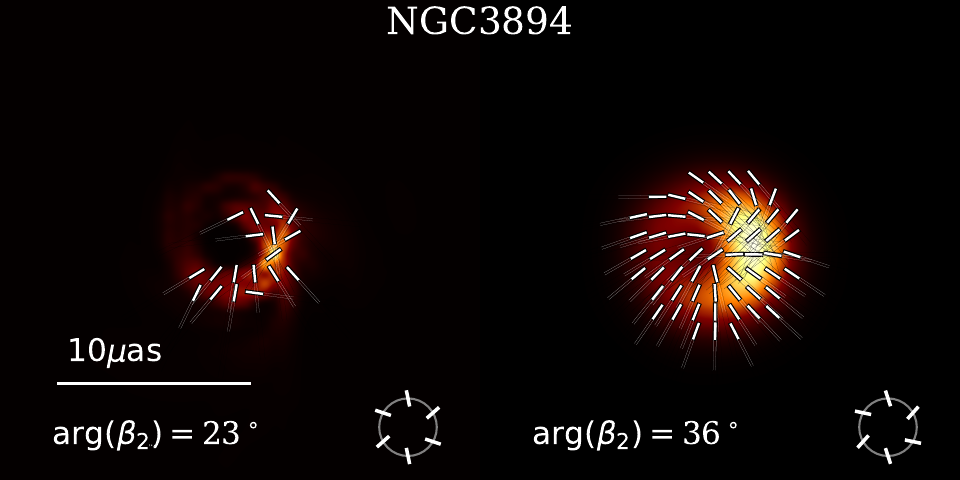}
\end{minipage}
\end{tabular}

\vspace{-0.08em}

\begin{tabular}{cc}
\centering
\begin{minipage}{0.4\textwidth}
    \centering
    \includegraphics[width=\linewidth]{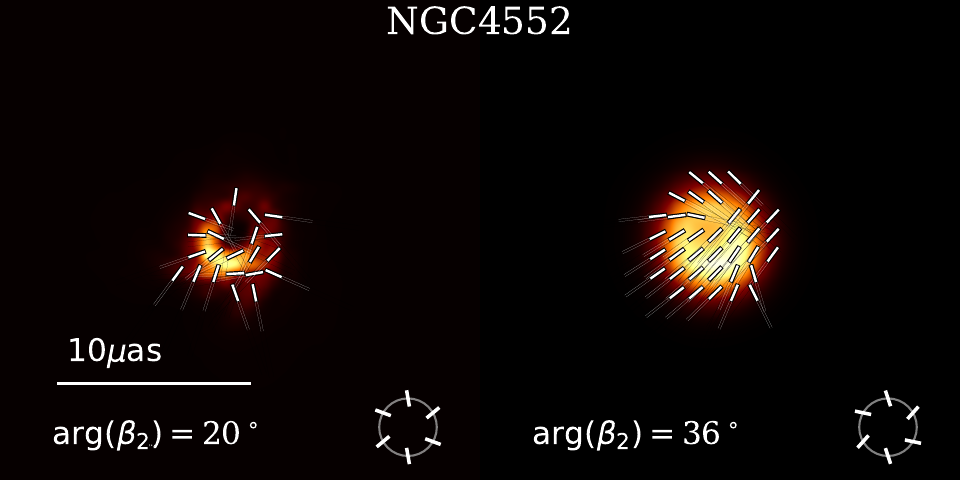}
\end{minipage}
\begin{minipage}{0.4\textwidth}
    \centering
    \includegraphics[width=\linewidth]{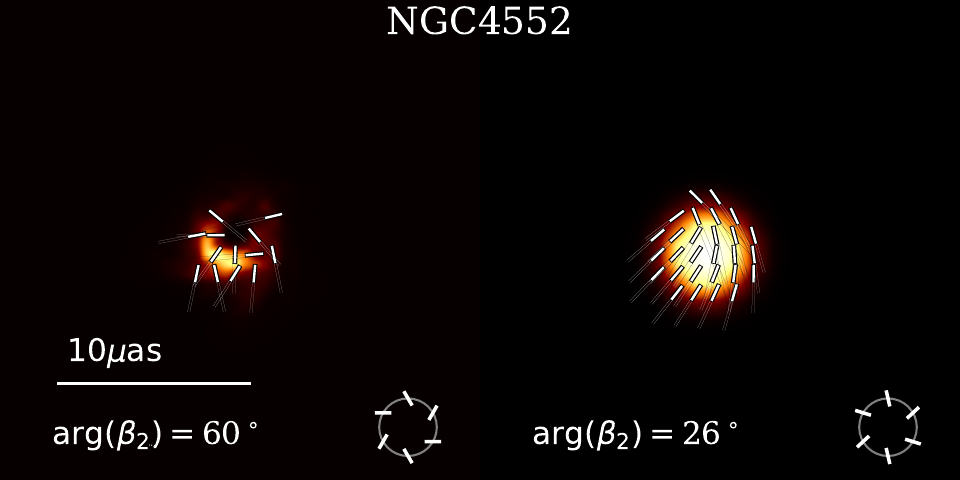}
\end{minipage}
\end{tabular}

\vspace{-0.08em}

\begin{tabular}{cc}
\centering
\begin{minipage}{0.4\textwidth}
    \centering
    \includegraphics[width=\linewidth]{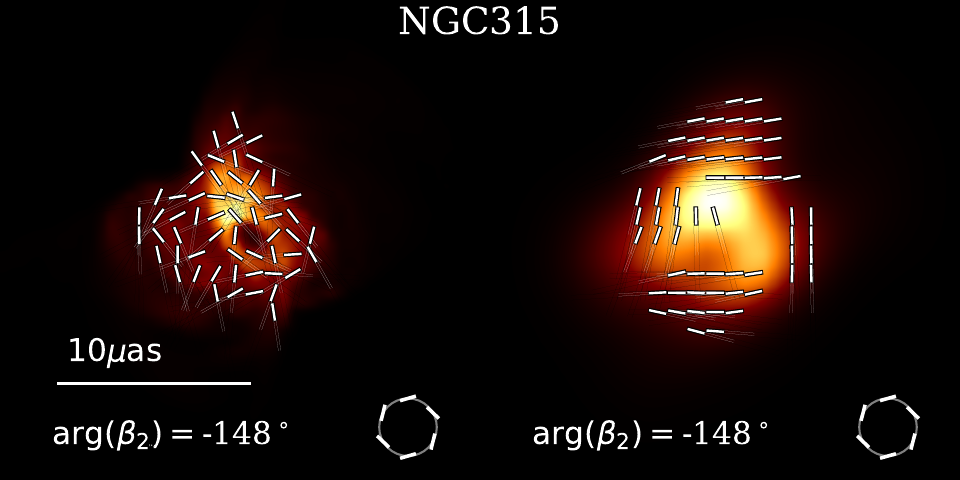}
\end{minipage}
\begin{minipage}{0.4\textwidth}
    \centering
    \includegraphics[width=\linewidth]{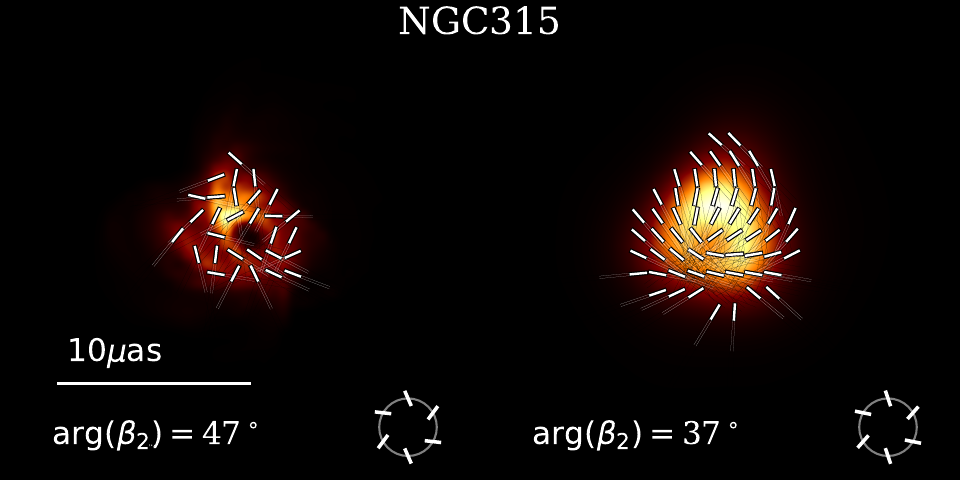}
\end{minipage}
\end{tabular}

\vspace{-0.08em}

\begin{tabular}{cc}
\centering
\begin{minipage}{0.4\textwidth}
    \centering
    \includegraphics[width=\linewidth]{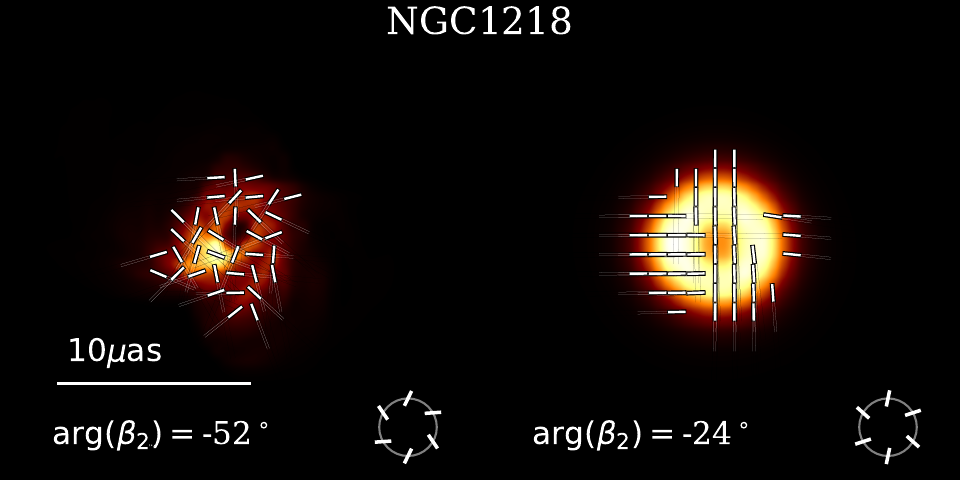}
\end{minipage}
\begin{minipage}{0.4\textwidth}
    \centering
    \includegraphics[width=\linewidth]{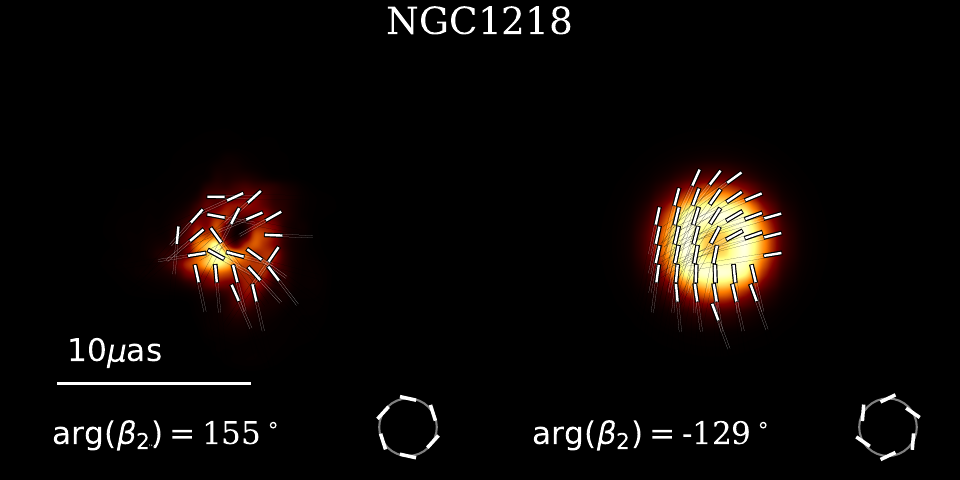}
\end{minipage}
\end{tabular}

\vspace{-0.08em}

\begin{tabular}{cc}
\centering
\begin{minipage}{0.4\textwidth}
    \centering
    \includegraphics[width=\linewidth]{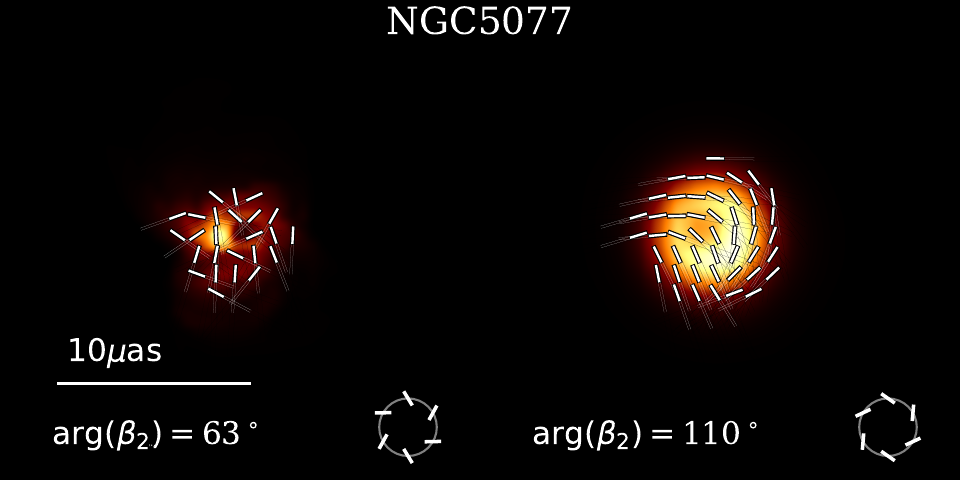}
\end{minipage}
\begin{minipage}{0.4\textwidth}
    \centering
    \includegraphics[width=\linewidth]{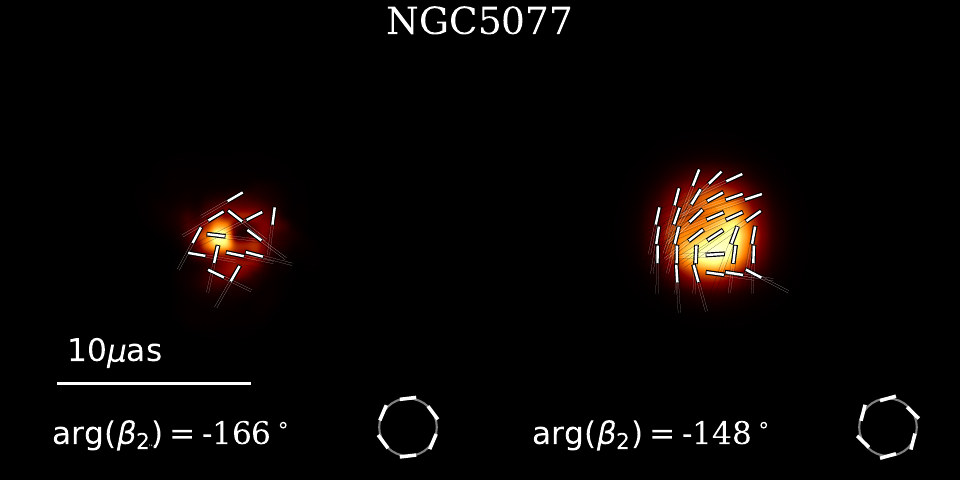}
\end{minipage}
\end{tabular}

\caption{---Continued.
}
\end{figure*}

\begin{figure*}[t]
\centering

\includegraphics[width=1\textwidth]{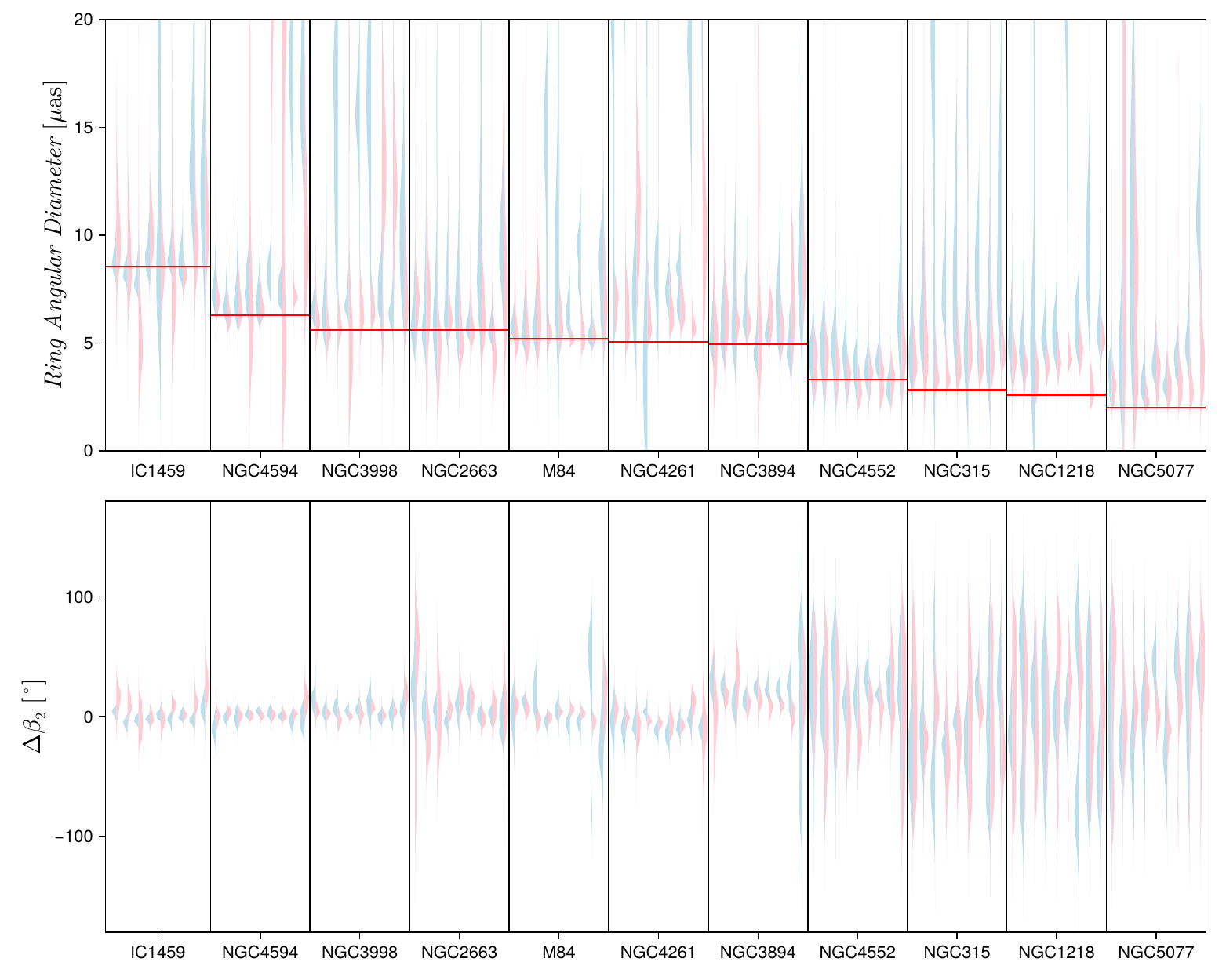}
\caption{
Posterior distributions of the ring angular diameter and $\Delta\beta_2$ for each frame of each source. The horizontal axis shows the frame index of synthetic observations for each target, and the vertical axis corresponds to the measured ring angular diameter and $\Delta\beta_2$. 
The posteriors from the 240\,GHz observations are shown in blue (left), while those from the 320\,GHz observations are shown in red (right). The red horizontal line indicates the shadow angular diameter for each source.
}
\label{fig:parameters}
\end{figure*}

In \autoref{fig:images}, we show the reconstructed images for frames with median values of the $(u,v)$ filling fraction metric \citep[see][for details]{Palumbo_2019}, along with the ground-truth GRMHD images for the 240 and 320\,GHz observations.
For most sources, both the source size and the polarization angle $\angle\beta_2$ are well reconstructed. In particular, for sources with shadow sizes larger than 5\,\uas, the overall morphologies of the models are well reproduced, including the central depression arising from the black hole shadow and the linear-polarization EVPA patterns.
These results demonstrate that even for the more complex GRMHD models, the detectability trends are broadly consistent with the thresholds derived in \autoref{subsec:measurement_thresholds}.

In fact, for sources with shadow sizes of $\gtrsim 3$\,\uas, the measured ring angular diameters show good agreement with the ground-truth shadow diameters, with a slight positive bias expected for measurements of the $n=0$ photon ring \citep[e.g.,][]{Kuramochi_2018, EHTC2017M87Paper6, EHTC2017SgrAPaper6}, as shown in the top panel of \autoref{fig:parameters}.
The median signal-to-noise ratios (SNRs) of the ring angular diameter measurements are 5.6 and 5.8 at 240 and 320\,GHz, respectively, across eight GRMHD models, excluding three sources (NGC\,315, NGC\,1218, and NGC\,5077) with shadow diameters smaller than 3\,\uas.
We note that some frames exhibit broader posterior distributions across multiple sources.
This behavior can be attributed either to (a) degraded $(u,v)$ coverage resulting from unfavorable weather conditions, or to (b) specific GRMHD realizations in which the ring emission is less dominant relative to jet emission or is not well closed due to significant asymmetry.
Both effects can hinder robust convergence of the diameter estimates.
These results highlight the importance of multiple observations to obtain reliable measurements of ring diameters that do not depend on a particular realization of weather conditions and time-variable emission structure.

For seven sources in which the magnetic-field structure is sufficiently detectable, the posterior distribution of $\Delta\beta_2$ is strongly concentrated near $0^\circ$, with root-mean-square deviations of approximately $12.5^\circ$ at 240\,GHz and $9.7^\circ$ at 320\,GHz (corresponding to EVPA uncertainties of $6.3^\circ$ and $4.9^\circ$, respectively) across these sources.
This demonstrates that the pitch angles of spiral EVPA patterns can be recovered with high precision for individual frames. Furthermore, these EVPA uncertainties are equivalent to a $1\sigma$ sensitivity to the rotation measure (RM) between 240 and 320\,GHz of approximately $\mathrm{RM} = 2.0 \times 10^{5}~\mathrm{rad\,m^{-2}}$, using standard error propagation.
This demonstrates that the pitch angles of spiral EVPA patterns are well measured for each frame. 
In contrast, four sources that do not meet this threshold (i.e., NGC\,315, NGC\,1218, NGC\,5077, and NGC\,4552) exhibit substantially larger deviations, owing to insufficient angular resolution to resolve the spatial EVPA structure of the ring emission.

\begin{figure*}[t]
\centering
\includegraphics[width=0.7\textwidth]{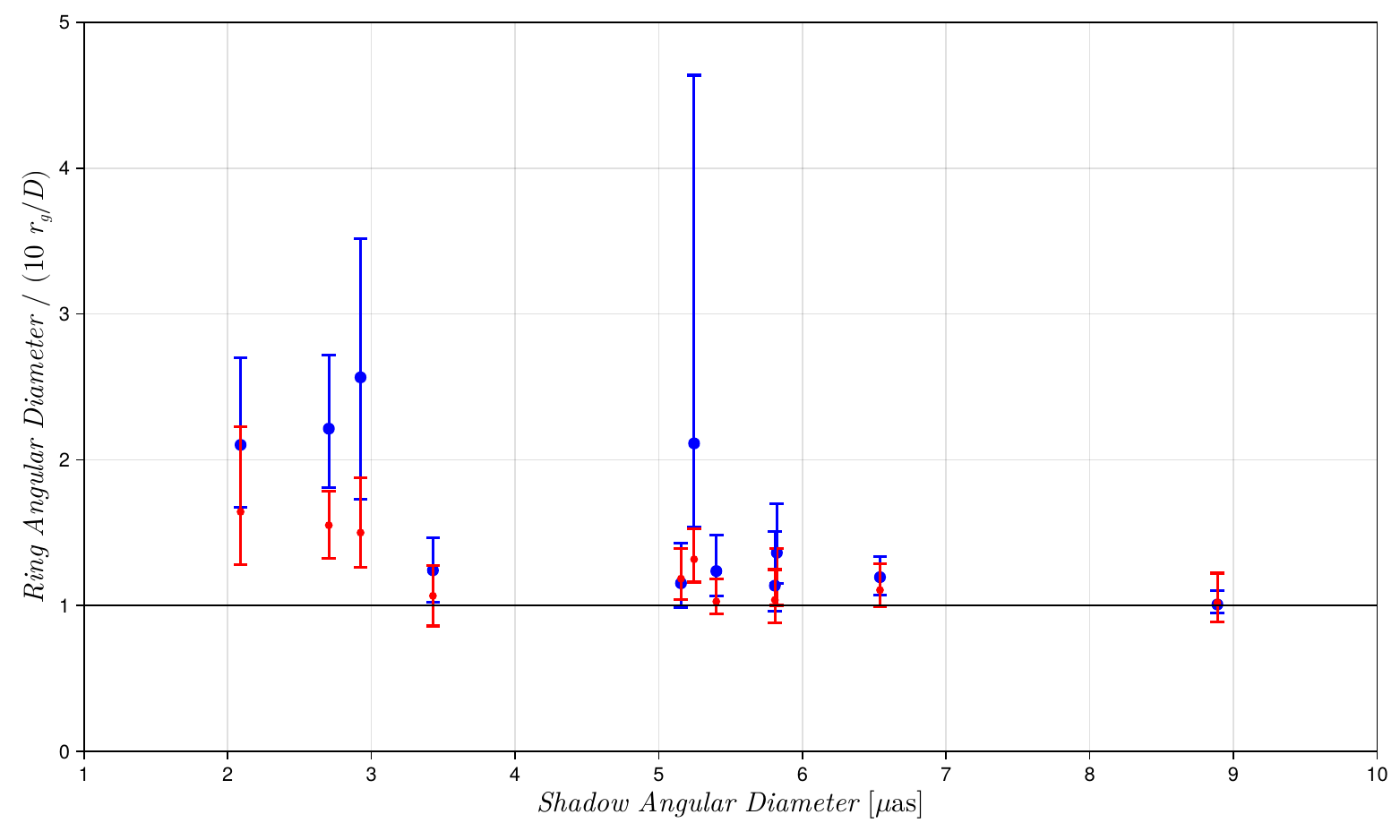}
\caption{
Shadow angular diameter dependence of the posterior probabilities for the median of the ring angular diameters measured through nine epochs of observations. 
The horizontal axis shows the shadow angular diameter $(10r_g/D)$, while the vertical axis displays the ring angular diameter normalized by the shadow angular diameter. 
Each point represents the median of the posterior distribution for each source, and the error bars denote the 95\,\% credible intervals.
}
\label{fig:parameters_vs_shadowsize}
\end{figure*}

In \autoref{fig:parameters_vs_shadowsize}, we present the posterior distributions of the medians of the ring angular diameters, normalized by their ground-truth shadow angular diameters, together with the residual $\beta_2$, obtained from nine epochs of synthetic observations, as a function of the ground-truth shadow angular diameter.
For shadow angular diameters larger than 3\,\uas, the inferred ring angular diameters are broadly consistent with the corresponding shadow diameters, with a median bias factor of $1.2$ and $1.1$, at 240 and 320\,GHz, respectively, broadly consistent with or slightly larger than in EHT observations of \m87 and \sgra of $\sim 1.1$ \citep[e.g.,][]{EHTC2017M87Paper6, EHTC2017SgrAPaper6}.
After correcting for this positive bias, the median ring angular diameters agree with the shadow angular diameters to within the $1\sigma$ accuracy of $7.8$\,\% and $7.6$\,\%, at each of two frequency bands. 
These results demonstrate that BHEX can measure the masses of SMBHs with the shadow angular diameter larger than 3\,\uas with an accuracy of $\lesssim 10$\,\% with the current simulation-based standard $\alpha$ calibration methods of the bias factor adopted in EHT observations of \m87 and \sgra \citep[e.g.,][]{EHTC2017M87Paper6, EHTC2017SgrAPaper6}.

\section{Conclusion}\label{sec:conclusion}
In this work, we evaluated the population of nearby horizon-scale targets accessible with a next-generation millimeter/submillimeter VLBI mission such as BHEX.
Using a suite of simulated observations, we quantified the number of nearby SMBHs for which proxies for black hole mass, magnetic field configuration, and black hole shadow can be robustly measured.
We further performed detailed imaging simulations based on dedicated GRMHD models for eleven nearby SMBHs to validate these population-level estimates.
Our main conclusions are summarized as follows:
\begin{itemize}
\item Using the latest SMBH number density model from \citet{Pesce_2021}, we find that BHEX could infer black hole masses for $\sim$70--90 sources from size measurements, constrain magnetic field structures via linear-polarization imaging for $\sim$20--30 sources, and resolve black hole shadows for $\sim$20--25 sources.
\item 
Targeted observations of real, nearby SMBHs ($\sim$50 sources) in the ETHER database are expected to yield measurements of $\sim$30 source sizes and $\sim$10 shadows and linear-polarization patterns. The detections of the mass proxy, based on the relatively easy size measurement, increase almost linearly for the first $\sim$20 observed sources, and anticipate a new detection for every $2$--$3$ additional targets until 50 observed sources. The other two proxies saturate earlier, as they require larger and brighter targets.
\item Detailed GRMHD imaging simulations show that for targets with shadow angular diameters larger than 5\,\uas, the overall source morphology is well recovered, including the central brightness depression associated with the black hole shadow and the characteristic EVPA patterns.
\item The median estimator of the ring angular diameter derived from nine epochs of synthetic observations agrees well with the ground-truth shadow diameter for targets with shadow sizes larger than 3\,\uas, with a positive bias of $10-20$\,\%, broadly consistent with EHT observations of \m87 and \sgra. After correcting for this bias, the inferred ring diameters are consistent with the shadow diameters within $1\sigma$ uncertainties of $\sim 8$\,\%. These results demonstrate that BHEX can measure the masses of SMBHs with the shadow angular diameter larger than 3\,\uas with an accuracy of $\lesssim 10$\,\% with the standard methods adopted in EHT observations of \m87 and \sgra.
\item For sources exceeding the measurement threshold of the magnetic field proxy, the posterior distributions of $\beta_2$ are typically consistent with the ground truth within $12.5^\circ$ and $9.7^\circ$, at 240 and 320\,GHz, respectively (corresponding to EVPA uncertainties of $6.3^\circ$ and $4.9^\circ$). This demonstrates that the pitch angles of spiral EVPA patterns can be robustly recovered.
\item In some frames, accurate ring diameter measurements are hindered by unfavorable weather conditions or by specific GRMHD realizations in which the ring morphology is weak or highly asymmetric. These results highlight the importance of repeated observations to mitigate the effects of atmospheric conditions and intrinsic source variability.
\end{itemize}

Overall, our results show that BHEX, or a comparable next-generation ground–space VLBI mission, would offer a powerful means of characterizing SMBH populations across a wide range of accretion states, radio-loudness regimes, host-galaxy environments, and viewing geometries. Our findings stand in contrast to the more pessimistic claims of \citet{BenZineb_2024}, demonstrating that the heuristic metrics used in that work are not an adequate substitute for end-to-end simulations when assessing the observational performance of BHEX-like ground–space VLBI arrays.

We note that this work presents only a glimpse of the total scientific capabilities attainable, focusing on a single-frequency, snapshot imaging analysis for synthetic observations. The scientific outcome will be further enhanced by advanced, physics-informed imaging and analysis techniques, including multi-frequency synthesis \citep[e.g.,][]{Arras_2022, Chael_2023}, dynamical imaging \citep[e.g.,][]{Johnson_2017, Bouman_2017, Arras_2022}, and direct emission modeling approaches \citep[e.g.,][]{Palumbo_2022, Chang_2024}. We leave such analyses for future work.

\begin{acknowledgments}
We appreciate the anonymous referee's constructive feedback and suggestions on the manuscript.
This work is financially supported by 
JST SPRING Grant Number JPMJSP2111, %
the Gordon and Betty Moore Foundation (GBMF-12987, GBMF-5278, GBMF-10423), %
the National Science Foundation (NSF; AST-1935980, AST-2034306, AST-2535855), %
the ULVAC-Hayashi Seed Fund from the MIT-Japan Program at MIT International Science and Technology Initiatives (MISTI) %
and the MIT Undergraduate Research Opportunity Program (UROP). %
The authors
acknowledge financial support from
MEXT/JSPS Grants-in-Aid for Scientific Research (KAKENHI) Grants (18H03721, 20H05860),
NSF (AST-2107681, AST-2132700, AST-2307887, OMA-2029670), %
and NINS Astrobiology Center program research (Grant Number AB0613). %

In addition to the above grants, technical and concept studies for BHEX have been supported by the Smithsonian Astrophysical Observatory, the Gordon and Betty Moore Foundation, the John Templeton Foundation, the Internal Research and Development (IRAD) program at NASA Goddard Space Flight Center, the University of Arizona, and the JAXA/ISAS Space Science Committee. BHEX is supported by initial funding from Fred Ehrsam. BHEX is funded in part by generous support from Mr. Michael Tuteur and Amy Tuteur, MD. We are grateful for advice and support from Ivan Selin. 

The Black Hole Initiative at Harvard
University is funded by grants from the John Templeton Foundation and the Gordon and Betty Moore Foundation to Harvard University.
\end{acknowledgments}

\begin{contribution}

We summarize the author contributions based on the Contributor Role Taxonomy (CRediT) as follows:
YA: Investigation, Methodology, Visualization, Writing -- original draft;
KA: Conceptualization, Funding acquisition, Methodology, Resources, Project administration, Software, Supervision, Writing -- original draft, Writing -- review \& editing;
DWP: Investigation, Methodology, Software, Writing -- original draft;
DCMP: Data curation, Methodology, Software;
AR: Conceptualization, Data curation, Methodology, Writing -- review \& editing;
PT: Data curation, Methodology, Software;
MNM: Investigation;
HY: Investigation;
AEB: Software;
AEH: Funding acquisition, Supervision;
SI: Data curation, Methodology;
NMN: Data curation;
KN: Funding acquisition, Supervision;
VR: Data curation;
XAZ: Data curation.
\end{contribution}

\software{
astropy \citep{astropy:2013, astropy:2018, astropy:2022}, \comrade \citep{Tiede_2022},
matplotlib \citep{Hunter:2007},
numpy \citep{harris2020array},
scipy \citep{2020SciPy},
\ehtim \citep{Chael_2016, Chael_2018},
\ngehtsim \citep{Pesce_2024_ngehtsim}
}

\appendix
\section{Estimating measurement uncertainties using Fisher information}\label{app:fisher_information}

We can approximate an arbitrary posterior distribution $p$ by a Gaussian one that takes the form

\begin{equation}
p = \frac{1}{\sqrt{\text{det}\left( 2 \pi \mathbf{M}^{-1} \right)}} \exp\left( - \frac{1}{2} \big[ \boldsymbol{\theta} - \boldsymbol{\mu_{\theta}} \big]^{\top} \mathbf{M} \big[ \boldsymbol{\theta} - \boldsymbol{\mu_{\theta}} \big] \right) ,
\end{equation}

\noindent where $\boldsymbol{\theta}$ is a parameter vector, $\boldsymbol{\mu_{\theta}}$ is a vector of the mean parameter values, and $\mathbf{M}$ is the Fisher information matrix.  The elements of the Fisher information matrix are given by the second derivative of the log-probability $\mathcal{P} = \ln(p)$ with respect to each parameter,

\begin{equation}
M_{ij} = -\left\langle \frac{\partial^2 \mathcal{P}}{\partial \theta_i \partial \theta_j} \right\rangle , \label{eqn:FisherMatrixElements}
\end{equation}

\noindent where the angle brackets denote an expected value.  For a Gaussian likelihood function, known measurement uncertainties, and uninformative (i.e., flat, wide) priors, the complex visibility log-posterior will be

\begin{equation}
\mathcal{P} = - \frac{1}{2} \sum_k \frac{| V_k - \hat{V}_k |^2}{\sigma_k^2} ,
\end{equation}

\noindent up to an overall constant term that is not important for our analysis.  Here, $V_k$ are the model data values, $\hat{V}_k$ are the measured data values, $\sigma_k$ are the measurement uncertainties, and the sum is taken over all data points $k$.  Given this form for $\mathcal{P}$, we can write \autoref{eqn:FisherMatrixElements} as

\begin{equation}
M_{ij} = \sum_k \frac{1}{2 \sigma_k^2} \left( \frac{\partial V_k}{\partial \theta_i} \frac{\partial V_k^*}{\partial \theta_j} + \frac{\partial V_k^*}{\partial \theta_i} \frac{\partial V_k}{\partial \theta_j} \right) ,
\end{equation}

\noindent where an asterisk denotes complex conjugation. \\

The inverse of the Fisher matrix corresponds to the covariance matrix for the effective Gaussian approximation to the target probability distribution, 

\begin{equation}
\boldsymbol{\Sigma} = \textbf{M}^{-1} .
\end{equation}

\noindent So the marginal distribution for the $n$th parameter can be approximated as a one-dimensional Gaussian with a variance of $\Sigma_{nn}$. \\

\noindent The analysis detailed in \autoref{sec:Populations_of_SMBHs} makes use of the \texttt{ngEHTforecast} package\footnote{\url{https://github.com/aeb/ngEHTforecast}}, which estimates the measurement precision for model parameters of interest using the Fisher information formalism described above.  The \texttt{ngEHTforecast} package implements the polarized ring model specified in \citet{Pesce_2022}, which includes model parameters describing the source structure as well as parameters describing the complex station gains.  Note that this approach avoids carrying out explicit fits of the model to synthetic data, instead assuming that the mean parameter vector $\boldsymbol{\mu_{\theta}}$ is equal to the input model parameter values (and then estimating the measurement uncertainty about that mean).

\end{document}